\documentclass[twocolumn, trackchanges]{aastex702}

\usepackage{hyperref}

\begin{document}





\title{Magnetic Activity as a Driver of Low-gravity-like Spectral Morphology in Late-M Dwarfs
}

\submitjournal{The Astronomical Journal}
\accepted{2026 October 20}

\author[0000-0001-5989-7594]{Chia--Lung Lin}
\affiliation{Steward Observatory, The University of Arizona, 933 N. Cherry Avenue, Tucson, AZ 85721, USA}
\email[show]{chialunglin@arizona.edu}

\author[orcid=0000-0003-3714-5855]{D\'aniel Apai}
\affiliation{Steward Observatory, The University of Arizona, 933 N. Cherry Avenue, Tucson, AZ 85721, USA}
\affiliation{Lunar and Planetary Laboratory, University of Arizona, 1629 E. University Boulevard, Tucson, AZ 85721, USA}
\affiliation{Alien Earths Team, NASA ICAR/NExSS, USA}
\email{apai@arizona.edu}

\author[orcid=0000-0002-3627-1676]{Benjamin V. Rackham}
\affiliation{Department of Earth, Atmospheric and Planetary Science, Massachusetts Institute of Technology, 77 Massachusetts Avenue, Cambridge, MA 02139, USA}
\affiliation{Kavli Institute for Astrophysics and Space Research, Massachusetts Institute of Technology, Cambridge, MA 02139, USA}
\email[]{brackham@mit.edu}

\author[orcid=0000-0002-8052-3893]{Prajwal Niraula}
\affiliation{Department of Earth, Atmospheric and Planetary Science, Massachusetts Institute of Technology, 77 Massachusetts Avenue, Cambridge, MA 02139, USA}
\affiliation{Kavli Institute for Astrophysics and Space Research, Massachusetts Institute of Technology, Cambridge, MA 02139, USA}
\email{pniraula@mit.edu}

\author[0000-0002-2132-5264]{Mark S. Giampapa}
\affiliation{Steward Observatory, The University of Arizona, Tucson, AZ 85721, USA}
\affiliation{Lunar and Planetary Laboratory, The University of Arizona, Tucson, AZ 85721, USA}
\affiliation{National Solar Observatory, 950 N. Cherry Avenue, Tucson, AZ 85719, USA}
\email{giampapa@arizona.edu}

\author[0000-0002-5322-2315]{Ana Glidden}
\altaffiliation{NSF Astronomy and Astrophysics Postdoctoral Fellow}
\affiliation{Center for Astrophysics, Harvard \& Smithsonian, 60 Garden Street, Cambridge, MA 02138, USA}
\email{ana.glidden@cfa.harvard.edu}

\author[orcid=0000-0003-2415-2191]{Julien de Wit}
\affiliation{Department of Earth, Atmospheric and Planetary Science, Massachusetts Institute of Technology, 77 Massachusetts Avenue, Cambridge, MA 02139, USA}
\email{jdewit@mit.edu}

\author[orcid=0009-0009-3020-3435]{Valeriy Vasilyev}
\affiliation{Max Planck Institute for Solar System Research, Justus-von-Liebig-Weg 3, 37077 G¨ottingen, Germany}
\email{vasilyev@mps.mpg.de}

\author[orcid=0000-0002-6087-3271]{Nadiia Kostogryz}
\affiliation{Max Planck Institute for Solar System Research, Justus-von-Liebig-Weg 3, 37077 G¨ottingen, Germany}
\email{kostogryz@mps.mpg.de}

\author[orcid=0000-0002-8842-5403]{Alexander Shapiro}
\affiliation{Max Planck Institute for Solar System Research, Justus-von-Liebig-Weg 3, 37077 G¨ottingen, Germany}
\affiliation{Institute of Physics, University of Graz, 8010 Graz, Austria}
\email{shapiroa@mps.mpg.de}










\begin{abstract}
We present a systematic investigation of the connection between magnetic activity and the low-surface-gravity-like near-infrared spectral morphology of late-M dwarfs.
Our sample includes TRAPPIST-1 observed with JWST NIRISS/SOSS and 22 late-M dwarfs observed with IRTF/SpeX, together with flare observations from K2, TESS, JWST NIRISS/SOSS, and literature-reported X-ray luminosities.
Using one-component PHOENIX/NewEra spectral modeling, we find that all targets are better reproduced by lower-surface-gravity model spectra than by models with the expected high gravities of field late-M dwarfs.
The retrieved surface gravities moderately correlate with independently measured gravity-sensitive spectral indices, with the strongest trends seen for FeH$_z$ at 0.998~$\mu$m and FeH$_J$ at 1.2~$\mu$m, both showing statistically significant Pearson and Spearman correlations with $p<0.005$.
They also show strong anti-correlations with flare luminosity and X-ray luminosity: for both relations, Pearson and Spearman coefficients are $<-0.9$, with $p \leq 1.0\times10^{-4}$, indicating that more active stars tend to have spectra favoring lower-gravity morphologies.
Time-series spectra of TRAPPIST-1 during two large flares show no clear short-timescale variations in selected photospheric indices, suggesting that persistent small-scale magnetic activity may play a more important role than large flares in modifying the photospheric temperature--pressure structure. 
A simplified radiative-transfer experiment further shows that activity-associated heating can produce lower-gravity-like spectral features even when the input surface gravity is fixed.
Our results suggest that underestimated surface gravities in late-M dwarfs may reflect atmospheres modified by frequent, small-scale magnetic activity, such as microflares, producing emergent spectra that mimic lower-gravity morphology.


\end{abstract}



\section{Introduction} \label{sec:intro}




Late-M dwarfs are important targets for stellar and exoplanetary studies because their small radii and low effective temperatures make them favorable hosts for atmospheric characterization of small planets \citep[e.g.,][]{Segura2005, Kaltenegger2009, 2017ApJ...850..121M, 2019AJ....158...27L, Wunderlich2019, TRAPPIST-1JWSTCommunityInitiative2024}. 
However, interpreting the spectra of planets around late-M dwarfs requires an accurate understanding of the stellar spectra. 
This remains challenging because late-M dwarf spectra are shaped by dense molecular absorption bands, pressure-sensitive features, and surface heterogeneity associated with spots, faculae, and magnetically active regions, making accurate modeling of their spectra challenging \citep[e.g.,][]{2013ApJ...772...79A,2018A&A...620A.180R,2018ApJ...853..122R, Rackham2023, Iyer2020, 2024AJ....168...82R, 2025AJ....170..297L}.

Previous studies have attempted to model the optical-to-near-infrared spectra of the late-M dwarf TRAPPIST-1 \citep{Gillon2016, 2017Natur.542..456G} as observed by HST \citep{Zhang2018, Wakeford2019, Garcia2022} or JWST \citep[e.g.,][]{2023ApJ...955L..22L, 2024ApJ...970L...4D, 2025ApJ...989..181P, 2025ApJ...979L...5R, Rathcke2025} using stellar atmosphere models such as PHOENIX ACES \citep{2013A&A...553A...6H}, PHOENIX NewEra \citep{2025A&A...698A..47H}, and SPHINX \citep{2023ApJ...944...41I, Iyer2026}.
A recurring and puzzling result from these studies is that model atmospheres with surface gravities close to the expected values of $\log g \sim 5$ for field late-M dwarfs struggle to reproduce the observed spectra of TRAPPIST-1, in line with the low-surface-gravity classification of TRAPPIST-1 from ground-based spectra \citep{Gillon2016, Burgasser2017, Gonzales2019}. 
In most cases, better fits are obtained using model spectra with substantially lower---and unphysical---surface gravities. 
{Importantly, this preference for low surface gravity is not limited to one-component spectral modeling. 
For example, \citet{2025ApJ...989..181P} performed three-component modeling of the out-of-transit NIRSpec/PRISM spectrum of TRAPPIST-1, including photospheric, cool-spot, and hot-facula components, and still retrieved photospheric surface gravities ($\log g \sim4$) substantially below the expected value of $\log g \simeq 5.24$. 
This suggests that adding multiple temperature components alone does not resolve the preference for low surface gravity.}
This does not necessarily imply that TRAPPIST-1 has genuinely low surface gravity, but may point to missing or incomplete atmospheric physics in current stellar atmosphere models.

Several hypotheses have been proposed or discussed for this low-gravity-like spectral morphology, including limitations in molecular opacities and model atmosphere spectra \citep{2024ApJ...970L...4D}, overestimated pressure broadening \citep{2026ApJ...997L...8G}, and magnetic-activity-related atmospheric effects \citep{Gonzales2019}.
Magnetic activity is a particularly interesting possibility because mid- to late-M dwarfs can remain highly active, showing frequent flares and, in some cases, detectable X-ray emission \citep[e.g.,][]{2004A&ARv..12...71G,2017ApJ...841..124V, Howard2019, 2023ApJ...959...64H, 2025ApJ...994L..31H, 2024AJ....168..234L,2025ApJ...979L...5R}. 
If magnetic activity modifies the thermal or pressure structure of the stellar atmosphere, it may cause the emergent spectrum to show low-surface-gravity-like morphology in both the pseudo-continuum and gravity-sensitive spectral features, such as the FeH Wing--Ford band and alkali lines \citep[e.g.,][]{2013ApJ...772...79A,2025AJ....170..213D}.

However, the relationship between low-gravity-like spectral morphology and magnetic activity has not yet been systematically tested for late-M dwarfs. In particular, it remains unclear whether the surface gravities retrieved from near-infrared spectral fitting, independently measured gravity-sensitive spectral indices, and magnetic activity indicators, such as flare luminosity and X-ray luminosity, are empirically linked.
In this work, we investigate this possible link using a sample of 23 late-M dwarfs, including TRAPPIST-1 observed with JWST NIRISS/SOSS and 22 additional late-M dwarfs observed with IRTF/SpeX, together with flare measurements from TESS and K2 and literature-reported X-ray observations. 
We find empirical links between underestimated surface gravities and magnetic activity, and use a toy model to explore a possible interpretation of this connection.

The structure of this paper is as follows. In Section~\ref{sec:observation_and_data_reduction}, we describe the observations and data reduction for the JWST NIRISS/SOSS spectra of TRAPPIST-1 and the IRTF/SpeX spectra of the remaining late-M dwarfs in our sample.
In Section~\ref{sec:analysis}, we present our spectral modeling framework, gravity-sensitive spectral index measurements, and magnetic activity indicators based on flare and X-ray observations.
In Section~\ref{sec:results}, we report the retrieved stellar parameters and examine the empirical relationships between the retrieved surface gravities, gravity-sensitive spectral indices, and magnetic activity indicators.
In Section~\ref{sec:discussion}, we discuss the possible role of magnetic activity in producing low-gravity-like spectral morphology, including the analysis of TRAPPIST-1 spectra during flare events and a simplified radiative-transfer modeling experiment.
Finally, in Section~\ref{sec:conclusions}, we summarize our main conclusions.

\section{Observations and Data Reduction} \label{sec:observation_and_data_reduction}

We analyzed near-infrared spectra of 23 late-M dwarfs with spectral types ranging from M6 to M9.5 (Table~\ref{tab:sample_properties}). 
TRAPPIST-1 was observed with JWST/NIRISS in Single Object Slitless Spectroscopy (SOSS) time-series mode. 
No other star with $T\leq3000$\,K has been observed by JWST in this same mode\footnote{Based on the data tabulated on \href{https://www.stsci.edu/~nnikolov/TrExoLiSTS/JWST/trexolists.html}{TrExoLiSTS} \citep{Nikolov2022} as of 2026\,June\,26.}.
Accordingly, for the remaining 22 late-M dwarfs studied here, we use spectra obtained with the SpeX spectrograph \citep{2003PASP..115..362R} on the 3.2-m NASA Infrared Telescope Facility (IRTF) at Maunakea, Hawaii.

\subsection{NIRISS/SOSS Observations of TRAPPIST-1}
TRAPPIST-1 was observed by JWST/NIRISS in SOSS time-series mode as part of two programs: GTO~1201 \citep{2017jwst.prop.1201L} and GO~2589 \citep{2021jwst.prop.2589L}. 
Five visits targeting TRAPPIST-1~f were obtained under GTO~1201 between 2022 and 2023. 
Under GO~2589, two visits targeting TRAPPIST-1~b were obtained in 2022 as well as two visits targeting TRAPPIST-1~c in 2022 and 2023.
{All nine visits used the same primary NIRISS/SOSS observing configuration: the SUBSTRIP256 subarray, the NISRAPID readout pattern, and 18 groups per integration, corresponding to an integration time of approximately 104\,s. The visits differed primarily in the number of integrations and therefore in their total observing durations.}
In total, there are nine JWST/NIRISS SOSS visits of TRAPPIST-1, with a total observing duration of 35.49 hr. The summary of these observations can be found in Table~\ref{tab:excluded_integrations}.

The NIRISS/SOSS data of TRAPPIST-1 used in this study were reduced using the package \texttt{exoTEDRF}, an end-to-end pipeline for JWST exoplanet time-series observations \citep{2023Natur.614..670F, 2023MNRAS.524..835R, 2024JOSS....9.6898R, 2025ApJ...985L..10A}. 
The pipeline first processes the raw detector data by identifying saturated pixels and flagging the affected columns for subsequent reduction steps. Bias correction is then applied to each integration. The background is subtracted from each image using the SUBSTRIP256 SOSS zodiacal model provided by STScI \citep{2023PASP..135d8002R}, followed by a $1/f$ noise correction for each integration. 
The pipeline then identifies and corrects jumps, such as cosmic-ray hits, using a time-domain sigma-clipping algorithm with a 7$\sigma$ rejection threshold. Ramp fitting is subsequently performed at the integration level using the corrected group-level reads. Finally, after flat-field correction and bad-pixel identification and correction, one-dimensional spectra are extracted with a 30-pixel-wide box aperture and wavelength calibrated.

The TRAPPIST-1 NIRISS/SOSS spectra were absolutely flux-calibrated using a sensitivity function derived from NIRISS/SOSS observations of the A-type star BD+60-1753, obtained through the COM/NIRISS 1091 program (PI: Andre Martel) and the CAL/CROSS 1536 program (PI: Karl Gordon). The BD+60-1753 observations were reduced using the same pipeline and procedures described above. The flux-calibration sensitivity curve was constructed by comparing the mean extracted count-rate spectra observed at different epochs with the absolute flux spectrum of BD+60-1753 from the \texttt{CALSPEC} library \citep{2014PASP..126..711B, 2020AJ....160...21B, 2022AJ....164...10B, 2025AJ....169...40B}.

\subsection{SpeX Observations of Other M6--M9.5 Dwarfs}
The SpeX spectra of the other late-M dwarfs considered here were obtained as part of an effort to characterize targets for the SPECULOOS Transit Survey \citep{Sebastian2021} and originally presented in \citet{2025AJ....170..213D}.
In short, the spectra were collected in the short-wavelength cross-dispersed (SXD) mode, which covers wavelengths from $0.8$ to $2.4~\mu\mathrm{m}$. 
The observations were conducted using the $0\farcs3$ slit, providing a resolving power of $R {\sim} 2000$.
For each observing sequence, flat-field and arc-lamp spectra were acquired to calibrate the detector response and wavelength solution, respectively. 
The raw frames were processed with \texttt{Spextool} v4.1 \citep{2004PASP..116..362C}. 
Atmospheric absorption was corrected with the \texttt{xtellcor} routine \citep{2003PASP..115..389V}, using A0\,V telluric standards observed at comparable airmasses to the science targets. 
The same standard stars were also used for the absolute flux calibration of extracted spectra of the late-M dwarfs.
The IRTF/SpeX observations of our late-M dwarf sample are summarized in Table~\ref{tab:sample_properties}.



\section{Analysis} \label{sec:analysis}

\subsection{Mean Out-of-transit and Flare-free NIRISS/SOSS TRAPPIST-1 Spectrum}
\label{subsec:mean oot spectrum of Trappist-1}


{TRAPPIST-1 is known to exhibit time-variable stellar spectra associated with changes in its heterogeneous photosphere, with wavelength-dependent variations observed on both intra-visit and rotational timescales \citep[e.g.,][]{Rathcke2025, 2025ApJ...989..181P}. 
A detailed characterization of this spectral variability and the corresponding surface heterogeneity is beyond the scope of this work. 
Our goal here is to characterize the representative quiescent near-infrared spectral morphology of TRAPPIST-1. 
We therefore construct a single mean spectrum from all nine NIRISS/SOSS visits, averaging over the rotational phases and surface configurations sampled by the observations.}

We only use exposures in which TRAPPIST-1 is in a quiescent state and is not obscured by any transiting planets.
Six flares were detected during the nine NIRISS/SOSS visits of TRAPPIST-1 \citep{2023ApJ...959...64H, 2025ApJ...994L..31H, 2025ApJ...989L..53V}. 
To construct a mean spectrum that is as free from flare contamination as possible, we excluded spectra within the flare-affected intervals, which were identified by inspecting the H$\alpha$ flare profiles in the light curves.

We integrated the flux from $0.64$ to $0.67~\mu\mathrm{m}$ to generate the H$\alpha$ light curves for all visits. 
The H$\alpha$ light curves were then normalized, and flares were identified using the criteria from \citet{2024AJ....168..234L}.
Specifically, a flare was required to satisfy three conditions: a positive deviation in the normalized flux, a signal-to-noise ratio larger than 3 with respect to the standard deviation of the light curve, and at least three consecutive data points meeting the first two criteria. 
{For each identified flare, the flare start was defined as the second data point before the flare peak in the H$\alpha$ light curve, while the end was defined as the first data point during the decay phase that fell below the $1\sigma$ scatter level of the corresponding quiescent H$\alpha$ light curve.}

\begin{figure}
    \centering
    \includegraphics[width=\textwidth, height=0.240\textheight, keepaspectratio]{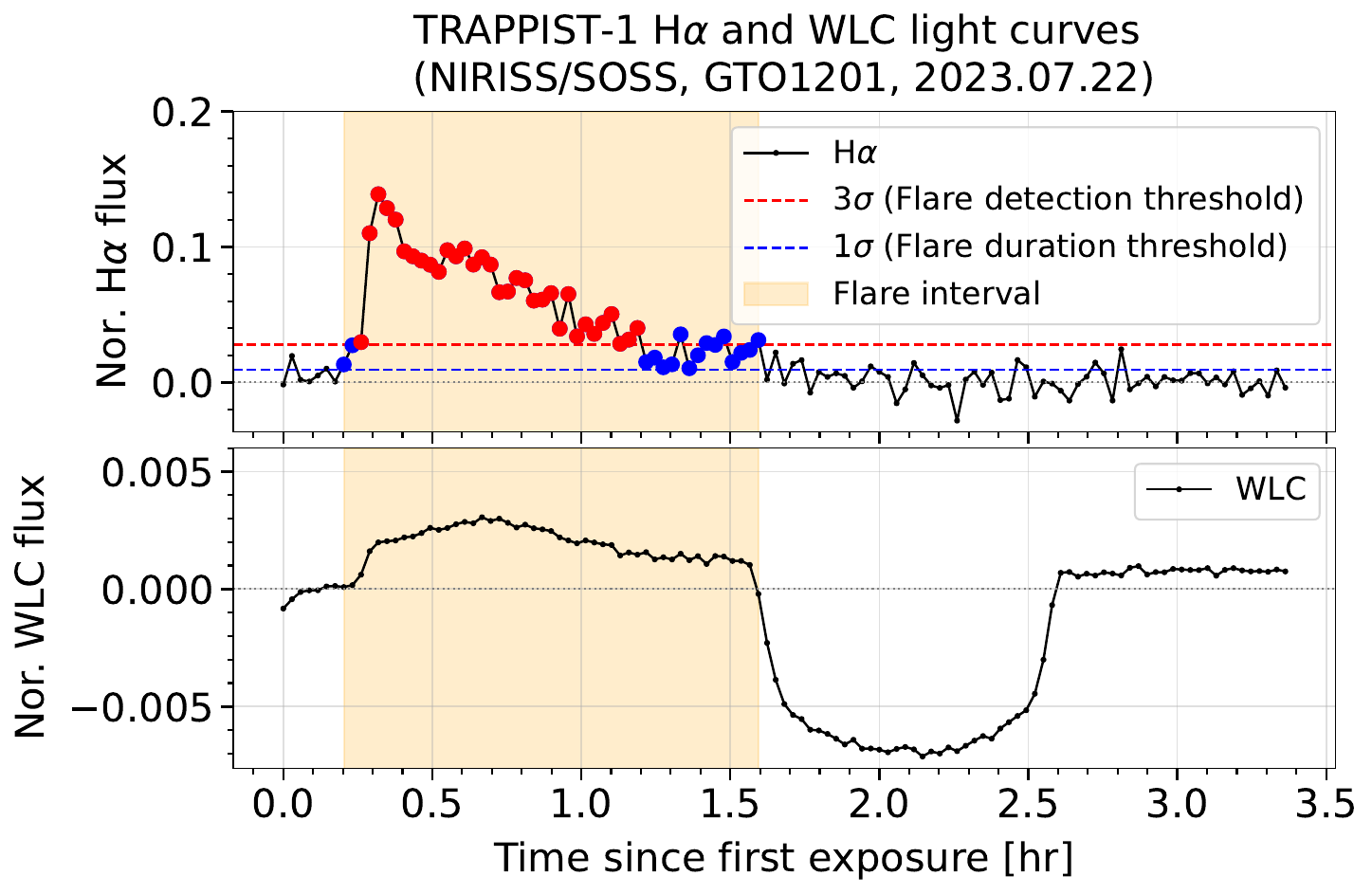}
    \caption{
     Example of the flare detection and duration criteria applied to the TRAPPIST-1 NIRISS/SOSS time-series observations, shown for the GTO~1201 visit on 2023 July 22. 
    The top panel shows the normalized H$\alpha$ light curve, and the bottom panel shows the corresponding normalized white-light curve (WLC). 
    The red and blue dashed lines in the top panel indicate the $3\sigma$ flare-detection threshold and the $1\sigma$ flare-duration threshold, respectively. 
    Red points mark measurements exceeding the $3\sigma$ detection threshold, while blue points mark measurements between the $1\sigma$ and $3\sigma$ thresholds used to trace the beginning and decay of the flare. 
    The shaded region indicates the adopted flare interval that is excluded when constructing the mean flare-free spectrum of TRAPPIST-1. 
    The flare interval is determined from the H$\alpha$ light curve, the WLC is shown for comparison.
    } 
    \label{fig:flare_detection_example}
\end{figure}

{Figure~\ref{fig:flare_detection_example} shows an example of this procedure for the GTO~1201 visit on 2023 July 22.
The flare interval is determined from the H$\alpha$ light curve using the $3\sigma$ detection threshold and the $1\sigma$ duration threshold, while the corresponding white-light curve is shown for comparison.
The shaded region marks the adopted flare interval that is excluded when constructing the mean flare-free spectrum of TRAPPIST-1.}

The shortest flare duration detectable with these criteria is approximately $5$--$8$ times the data sampling interval. 
These requirements were adopted to reduce false detections caused by noise fluctuations, following the detailed study of approximately one million TESS light curves by \citet{2024AJ....168..234L}.
As a result, we re-detected six flares in the visits on 2022 July 20, 2022 October 28, and 2023 October 31 from GO~2589, and on 2022 October 28, 2023 June 15, and 2023 July 22 from GTO~1201. 
All spectra within these flare intervals were excluded from the mean-spectrum estimate.

{For each visit, the integration intervals affected by planetary transits and/or flares were excluded before calculating the mean spectrum.
The excluded integration ranges and the corresponding masked fraction for each visit are listed in Table~\ref{tab:excluded_integrations}.
In total, 14.66 hr out of the 35.49 hr of observations were excluded, corresponding to 41.32\% of the full dataset.}
Finally, we obtained the mean spectrum of TRAPPIST-1 using all out-of-transit and flare-free time-series spectra from the nine NIRISS/SOSS visits. 
This mean spectrum was used for the subsequent analysis in this work.

\subsection{Modeling the Near-Infrared Stellar Spectra}
\label{subsec:modeling_NIR_spectra}
We use the \texttt{PHOENIX} NewEra stellar atmosphere models \citep{2025A&A...698A..47H} to fit the NIRISS/SOSS and SpeX spectra of the late-M dwarfs analyzed in this study.
The PHOENIX/NewEra model grid consists of synthetic stellar spectra with broad wavelength coverage, including the near-infrared, spanning effective temperatures from 2300 to 12,000 K in steps of 100 K, metallicities from $-4.0$ to $+0.5$ dex in steps of 0.5 dex, and surface gravities $\log g$ from 0.0 to 6.0 in steps of 0.5 dex.
To generate synthetic spectra at arbitrary stellar parameters within the model parameter space, we use \texttt{speclib} \citep{2024AJ....168...82R, speclibv0.1.0b11}  to perform grid interpolation for our spectral modeling analysis.

Accurate spectral modeling of late-M dwarfs at optical and near-infrared wavelengths remains challenging because these spectral regions are strongly affected by complex molecular absorption features and uncertainties in stellar model fidelity \citep[e.g.,][]{2018A&A...620A.180R, 2023ApJ...944...41I, Iyer2026, 2023ApJ...955L..22L}. 
Previous full-band spectral modeling of TRAPPIST-1 has also shown substantial model--data discrepancies at wavelengths shorter than 1.3~$\mu$m, particularly around 0.7--0.8~$\mu$m and 1.0--1.1~$\mu$m, even when multicomponent or low-$\log g$ PHOENIX models are considered \citep{2023ApJ...955L..22L,2025ApJ...989..181P}. 
Motivated by these short-wavelength discrepancies, and to apply a consistent fitting range across the NIRISS/SOSS and IRTF/SpeX samples, we restrict our spectral modeling to wavelengths longer than 1.3~$\mu$m.

We performed parameter estimation using dynamic nested sampling implemented in 
\texttt{dynesty} \citep{2020MNRAS.493.3132S}, adopting a configuration similar to that used by 
\citet{2026arXiv260324585N} in their Section~3.4. 
We used 1000 live points together with the \texttt{rslice} sampler and \texttt{multi} bounding option. 
The iteration limit was set to 500,000, which was chosen to give the sampler sufficient room to map the posterior distribution without imposing an early stopping constraint.

To capture residual scatter not described by the reported photon-noise uncertainties alone, we included an additional error-inflation term in the likelihood function.
The log-likelihood is given by
\begin{equation}
\ln \mathcal{L} =
-\frac{1}{2}
\sum_{\lambda}
\left[
\frac{\left(F_{\rm obs,\lambda}-F_{\rm model,\lambda}\right)^2}
{\sigma_{\lambda}^{2}}
+
\ln\left(2\pi\sigma_{\lambda}^{2}\right)
\right],
\end{equation}
where
\begin{equation}
\sigma_{\lambda}^{2}
=
\sigma_{\rm photon,\lambda}^{2}
+
\sigma_{\rm jitter,\lambda}^{2}.
\end{equation}
Following \citet{Narrett2024}, we parameterized the jitter term as
\begin{equation}
\sigma_{\rm jitter,\lambda}
=
f_{\rm var} F_{\rm model,\lambda},
\end{equation}
where $f_{\rm var}$ is a fitted coefficient that sets the amplitude of the extra noise component. 
This allows the additional uncertainty to increase with the modeled flux level, rather than being treated as a wavelength-independent constant.
{
Linear interpolation between stellar atmosphere grid points may introduce small residual spectral differences and reduce model fidelity \citep{2024AJ....168...82R}. 
We performed a test of this impact by leaving one grid point out for a representative late-M-dwarf model, which indicated that these differences are small for interpolation in $T\mathrm{eff}$ and $\log g$.
The fitted error-inflation term can partially account for this additional uncertainty in the likelihood.
}

Using this framework, we performed one-component spectral modeling for all targets, assuming that the observed spectrum is described by a single quiescent photosphere. We used the same parameter boundaries and prior choices across the full sample.
We considered two fitting scenarios. In the first scenario, the photospheric effective temperature ($T_{\rm phot}$), metallicity ([Fe/H]), and surface gravity ($\log g$) were all treated as free parameters. In the second scenario, $T_{\rm phot}$ and [Fe/H] were fitted using the same broad prior ranges as in the first scenario, while the allowed range of $\log g$ was restricted.
For the first scenario, we adopted uniform priors of $T_{\rm phot} \in [2300, 3500]$\,K, ${\rm [Fe/H]} \in [-1, 0.5]$, $\log g \in [2.5, 5.5]$, and $\ln f_{\rm var} \in [-7, 0]$.
For the second scenario, we restricted the surface gravity to $\log g \in [5.2, 5.3]$, a range consistent with the expected main-sequence surface gravities of the late-M dwarfs in our sample \citep[e.g.,][]{2013ApJS..208....9P}, while keeping the priors on the other parameters identical to those used in the first scenario.
{
For the stellar radius, we adopted truncated Gaussian priors for all targets in our sample, centered on the literature value of TRAPPIST-1 (see Table~\ref{tab:trappist1_adopted_parameters}), with the standard deviation set to 40\% of that value and truncation bounds of $[0.05,\,0.2]\,R_{\odot}$, given the similar spectral types of the targets.
}
This range was chosen to encompass the plausible radii of the objects in our sample.
For the distance, we adopted truncated Gaussian priors centered on the literature value for each target listed in Table~\ref{tab:sample_properties}, with boundaries set to $\pm 10\%$ of the corresponding value.
These settings are summarized in Table~\ref{tab:one_component_priors}.

\subsection{Measurements of Spectral Gravity-Sensitive Indices}
\label{subsec: measure spectral gravity sensitive indices}
Several surface-gravity-sensitive spectral features are present across the near-infrared, including the FeH Wing--Ford band at 0.998~$\mu$m (hereafter FeH$_z$), VO$_z$ at 1.058~$\mu$m, FeH$_J$ at 1.200~$\mu$m, K\,\textsc{i}$_J$ at 1.244~$\mu$m, and H-cont at 1.560~$\mu$m \citep{2013ApJ...772...79A}. 
We seek to investigate the correlation between the surface gravities retrieved from the NIR pseudo-continuum modeling of the free-$\log g$ scenario described in Section~\ref{subsec:modeling_NIR_spectra} and the indices inferred from these gravity-sensitive features.

Importantly, all of these indices except H-cont at 1.560~$\mu$m lie outside the wavelength range used in our IRTF/SpeX and NIRISS/SOSS spectral modeling retrievals. 
Because H-cont is included in the fitted spectrum, its behavior may contribute to the retrieved $\log g$ values. 
Therefore, the correlations between the retrieved $\log g$ values and the other gravity-sensitive indices provide a partially independent test of whether the behavior of H-cont within the fitted wavelength range is also reflected in gravity-sensitive spectral features outside this range.



We estimated the indices of these gravity-sensitive features using the method introduced by \citet{2013ApJ...772...79A}. The index is defined as
\begin{equation}
\label{eq:spectral_index}
{\rm index}
=
\left(
\frac{\lambda_{\rm line}-\lambda_{\rm cont1}}
{\lambda_{\rm cont2}-\lambda_{\rm cont1}}
F_{\rm cont2}
+
\frac{\lambda_{\rm cont2}-\lambda_{\rm line}}
{\lambda_{\rm cont2}-\lambda_{\rm cont1}}
F_{\rm cont1}
\right)
/F_{\rm line}.
\end{equation}
Here, $\lambda_{\rm line}$, $\lambda_{\rm cont1}$, and $\lambda_{\rm cont2}$ are the central wavelengths of the line, blue-continuum, and red-continuum bands, respectively. 
The quantities $F_{\rm line}$, $F_{\rm cont1}$, and $F_{\rm cont2}$ are the average fluxes measured within the corresponding bands.
The numerator gives the expected flux at $\lambda_{\rm line}$ in the absence of absorption or emission, based on a linear interpolation between the average fluxes measured in the two continuum bands.

Table~\ref{tab:gravity_index_bands} summarizes the band definitions and adopted bandwidths for the spectral indices measured in this study. 
These values were adopted directly from Table~4 of \citet{2013ApJ...772...79A}, except for H-cont, for which we followed the modified definition proposed by \citet{2025AJ....170..213D}. 
For the H-cont index, we adopted a blue-continuum band centered at 1.270~$\mu$m to accommodate the limited wavelength coverage of the IRTF/SpeX spectra after masking regions affected by strong telluric absorption. 
The adopted bandwidths are broader than one spectral resolution element in the IRTF/SpeX and NIRISS/SOSS spectra analyzed here. 
No interpolation or resampling of the observed spectra was applied, and the average flux was calculated directly from the spectral points falling within each band.

For the FeH$_z$, VO$_z$, FeH$_J$, and K~I$_J$ indices, a higher index value corresponds to a deeper absorption, higher-gravity-like feature.
The H-cont index instead measures the shape of the blue side of the $H$~band rather than the depth of an absorption feature. 
Higher H-cont values indicate a more triangular, low-gravity-like $H$-band continuum, whereas lower values indicate a less triangular, higher-gravity-like continuum.




\subsection{Magnetic Activity Indicators}


\subsubsection{Stellar Flare Observations and Re-analysis of TRAPPIST-1 K2 Short-cadence Flares}
TRAPPIST-1 is known to be a flare star with a high level of flare activity \citep[e.g.,][]{2017ApJ...841..124V,2023ApJ...959...64H,2026ApJ..1004L...3V}.
For the remaining late-M dwarfs in our sample, we identify the flare stars and retrieve their flare measurements from the flare-star catalog of \citet{2024AJ....168..234L}.
Among our late-M dwarf sample, 19 stars were analyzed by \citet{2024AJ....168..234L}, and flares were detected on 12 of them.

{
Of the three stars not included in \citet{2024AJ....168..234L}, VB~10 is a known flare star, with flares previously detected at UV, X-ray, and near-infrared wavelengths \citep{1999AJ....118.1369K,Linsky1995,Fleming2000,Kanodia2022}. However, no long-term photometric observations such as Kepler/K2 or TESS are available for this target, so we do not include VB~10 in our flare analysis. Its reported X-ray luminosity is still included in our X-ray activity analysis in the subsequent sections. 
For the other two stars, we did not find any published flare detections. \citet{2000AJ....120.1085G} reported a single-epoch non-detection of H$\alpha$ for 2MASSW~J0109216+294925, with no repeated spectroscopic monitoring reported for this target. 
SIPS~J1607-0442 was included in the photometric monitoring of \citet{2014AJ....147...94D} using the ground-based CTIO 0.9~m telescope, but those observations were obtained in the $I$~band with 600~s exposures and were not designed as a high-cadence flare survey.
Since M-dwarf flare contrast decreases toward redder wavelengths and flare detection becomes less sensitive in redder bands \citep[e.g.,][]{Chang2020,Maas2022,2025AJ....170..297L}, short-duration flares can also be diluted by long exposures. The absence of a reported flare therefore does not provide a strong constraint on the flare activity of this star. We consider the flare status of these two stars to be unclear based on the current literature. 
Therefore, including TRAPPIST-1, 13 late-M dwarfs in our sample have K2/TESS-based flare measurements used in the following analysis.
}





Although TRAPPIST-1 was also observed by TESS, its TESS light curve is strongly contaminated, and no flares were reported by \citet{2024AJ....168..234L}.
Therefore, to place TRAPPIST-1 on a TESS-equivalent flare-energy scale consistent with the other stars in our sample, we re-analyze its K2 short-cadence Campaign 12 observations, starting from the Target Pixel File (TPF), and convert the measured K2 flare energies into the TESS bandpass.
We extract the short-cadence light curve of TRAPPIST-1 using the \texttt{Lightkurve} package \citep{2018ascl.soft12013L}, following the procedure described by \citet[][Section~2]{2017ApJ...841..124V}. 
In brief, this procedure corrects the K2 pointing-related systematics by using the centroid motion of the target on the detector, which is important for reducing instrumental trends while preserving flare-like signals. 
We use the resulting short-cadence light curve for our flare identification and energy measurements.

We detrend the extracted K2 short-cadence light curve of TRAPPIST-1 and identify flare candidates following the procedure described by \citet[][Section~2.1]{2024AJ....168..234L}. 
After that, we normalize the light curve by dividing the flux excess, $\Delta F(t)$, by the median flux of the detrended light curve, $\widetilde{F}$,
\begin{equation}
  \frac{\Delta F(t)}{\widetilde{F}} = \frac{F(t)-\widetilde{F}}{\widetilde{F}}.
  \label{eq:eq_amplitude}
\end{equation}
We then estimate the equivalent duration in units of seconds for each flare candidate following \citet{1972Ap&SS..19...75G},
\begin{equation}
  ED = \int \frac{\Delta F(t)}{\widetilde F} dt,
  \label{eq:ED}
\end{equation}
which represents the amount of time required for the star in quiescence to emit the same amount of energy as the cumulative flare energy.
For each flare candidate, we determine the four features defined by \citet[][Section~2.2]{2024AJ....168..234L}: the flare duration, $\delta t$, in days; the impulsive-to-decay time ratio, $\delta t_1 / \delta t_2$; the upper-to-lower equivalent duration ratio, $ED_1 / ED_2$; and the peak-amplitude signal-to-noise ratio, $A_f / \sigma(F)$. 
We adopt $\sigma(F)=0.0052$ from the detrended K2 short-cadence light curve of TRAPPIST-1.
We further apply the machine-learning-based classifier from \citet{2024AJ....168..234L} to distinguish true flares from the flare candidates. 
This process confirms 44 flares in the K2 short-cadence light curve, two more than the 42 flares reported by \citet{2017ApJ...841..124V}, and is consistent with the recent independent analysis of \citet{2026ApJ..1004L...3V}.


We estimate the energies of the flares detected on TRAPPIST-1 in the Kepler bandpass using
\begin{equation}
E_{f, \mathrm{Kepler}} = ED \times L_{*, \mathrm{Kepler}},
\label{eq:flare_energy}
\end{equation}
where $L_{*, \mathrm{Kepler}}$ is the quiescent luminosity of TRAPPIST-1 in the Kepler bandpass.
We adopt the Gaia XP spectrum of TRAPPIST-1 and convolve it with the Kepler bandpass response to obtain the quiescent flux of TRAPPIST-1 in the Kepler bandpass, $F_{*, \mathrm{Kepler}} = 2.24 \times 10^{-12}~\mathrm{erg~s^{-1}~cm^{-2}}$.
Using the distance to TRAPPIST-1, $d = 12.467~\mathrm{pc}$, and the relation
$L_{*, \mathrm{Kepler}} = 4\pi d^{2}F_{*, \mathrm{Kepler}},$
we obtain a quiescent Kepler-band luminosity of $L_{*, \mathrm{Kepler}} = 4.17 \times 10^{28}~\mathrm{erg~s^{-1}}$ for TRAPPIST-1, which we adopt in Equation~\ref{eq:flare_energy} to estimate flare energy.
We then convert the Kepler-band flare energies to flare energies in the TESS bandpass by adopting a flare temperature of 3500~K. This value reflects the “thermostat” effect reported for ultracool-dwarf flares in the near-IR: a substantial fraction of the flare energy goes toward molecular dissociation, which limits further atmospheric heating, thereby constraining increases in the effective flare continuum temperature \citep{2026ApJ...998L...7S}. 
Such behavior is consistent with recent JWST time-series spectroscopy of flares on TRAPPIST-1 \citep{2023ApJ...959...64H, 2025ApJ...994L..31H}.
The conversion is carried out using
\begin{equation}
E_{\mathrm{flare,TESS}} =
c_{\mathrm{TESS/Kepler}}(T_{\mathrm{flare}}=3500~\mathrm{K})
E_{\mathrm{flare,Kepler}},
\label{eq:Kepler_to_TESS_conversion}
\end{equation}
where $c_{\mathrm{TESS/Kepler}}(T_{\mathrm{flare}}=3500~\mathrm{K})$ is defined as the ratio between the flux of a 3500~K blackbody integrated over the TESS bandpass response function and that integrated over the Kepler bandpass response function. We obtain a conversion factor of $c_{\mathrm{TESS/Kepler}}(T_{\mathrm{flare}}=3500~\mathrm{K}) = 1.452$.

We convert the Kepler-band flare energies of TRAPPIST-1 into the TESS bandpass only to make them comparable with the TESS-based flare measurements of the other late-M dwarfs in our sample. We do not convert all flare energies into bolometric energies because this would require assuming a flare temperature distribution for all stars. 
Since it is still unclear whether different M dwarfs have the same flare temperature distribution, such a conversion could introduce additional uncertainty. 
Therefore, we keep the flare luminosity comparison in the TESS bandpass.
We also note that our flare-energy estimates do not include the small correction from the photon-weighted band-integrated signal measured by photon-counting instruments to the corresponding energy-flux convention. \citet{2026ApJ..1004L...3V} discuss this correction, which may introduce a small systematic offset at the few-percent level in the inferred flare energies if it is not included.
This is small compared with the order-of-magnitude range of flare luminosities in our sample, and therefore should not affect the main trends discussed here.



{We also estimate the TESS-band energies of the six flares detected in the JWST NIRISS/SOSS observations of TRAPPIST-1.
We first integrate the NIRISS/SOSS time-series spectra over the TESS bandpass to construct a TESS-band light curve for each visit. 
The flare intervals are defined from the H$\alpha$ light curves following the procedure described in Section~\ref{subsec:mean oot spectrum of Trappist-1}. 
We then normalize the TESS-band light curves and calculate the equivalent duration of each flare using Equation~\ref{eq:ED}. 
The TESS-band flare energies are estimated using Equation~\ref{eq:flare_energy}, adopting a quiescent TESS-band luminosity of $L_{\star,\rm TESS}\simeq3.2\times10^{29}$~erg~s$^{-1}$. 
The resulting energies of all the TRAPPIST-1 flares in this study are summarized in Table~\ref{tab:trappist1_flare_energies}. }

Finally, to account for the different observational baselines and compare the flare activity levels among our sample, we estimate the TESS-band flare luminosity, $L_{\rm flare, TESS}$, in units of $\mathrm{erg~s^{-1}}$ using
\begin{equation}
L_{\rm flare, TESS} = \frac{\sum E_{\rm flare,TESS}}{t_{\rm obs}},
\label{eq:tess_flare_luminosity}
\end{equation}
where $t_{\rm obs}$ is the total duration of the observations. 
The resulting $L_{\rm flare, TESS}$ values for our late-M dwarf sample are summarized in Table~\ref{tab:flare_xray_luminosities}.
 

\subsubsection{X-ray Observations}
We also collect the literature-reported X-ray luminosities of our sample of late-M dwarfs.
Coronal X-ray emission is also associated with magnetic activity, as it traces the magnetically heated hot plasma in the stellar corona \citep[e.g.,][]{2004A&ARv..12...71G, 2019A&A...628A..41P}. 
Therefore, X-ray luminosity provides a complementary indicator of the magnetic activity levels of these stars, in addition to the flare luminosity derived from the optical light curves of TESS and K2.

Among the 23 late-M dwarfs in our sample, X-ray luminosity measurements or upper limits are available for nine stars.
Six of them have reported X-ray detections: VB~10, TVLM~513-46, LP~412-31, DENIS-P~J1048-3956, LHS~2065, and TRAPPIST-1.
For three additional targets (LP~271-25, 2MASSI~J1835379+325954, and LP~944-20), only upper limits on their X-ray luminosities are available.

Four of the stars with detected flares also have detected X-ray luminosities: LP~412-31, DENIS-P~J1048-3956, LHS~2065, and TRAPPIST-1.
Their X-ray luminosities range from $(1.70\pm0.70)\times10^{25}~\mathrm{erg~s^{-1}}$ for DENIS-P~J1048-3956 to $(1.60\pm0.30)\times10^{27}~\mathrm{erg~s^{-1}}$ for LP~412-31.
The collected X-ray luminosities, together with their corresponding references, are summarized in Table~\ref{tab:flare_xray_luminosities}.

\section{Results} \label{sec:results}

\subsection{Stellar Spectral Modeling: Free-$\log g$ versus Restricted-$\log g$}
Table~\ref{tab:retrieved_params_free_fixed_logg} summarizes the best-retrieved stellar parameters from the one-component PHOENIX/NewEra spectral modeling of the late-M dwarfs in our sample for both the free-$\log g$ and restricted-$\log g$ scenarios. 
\subsubsection{{TRAPPIST-1 Retrieval from NIRISS/SOSS}}
\begin{figure}
    \centering
    \includegraphics[width=\textwidth, height=0.35\textheight, keepaspectratio]{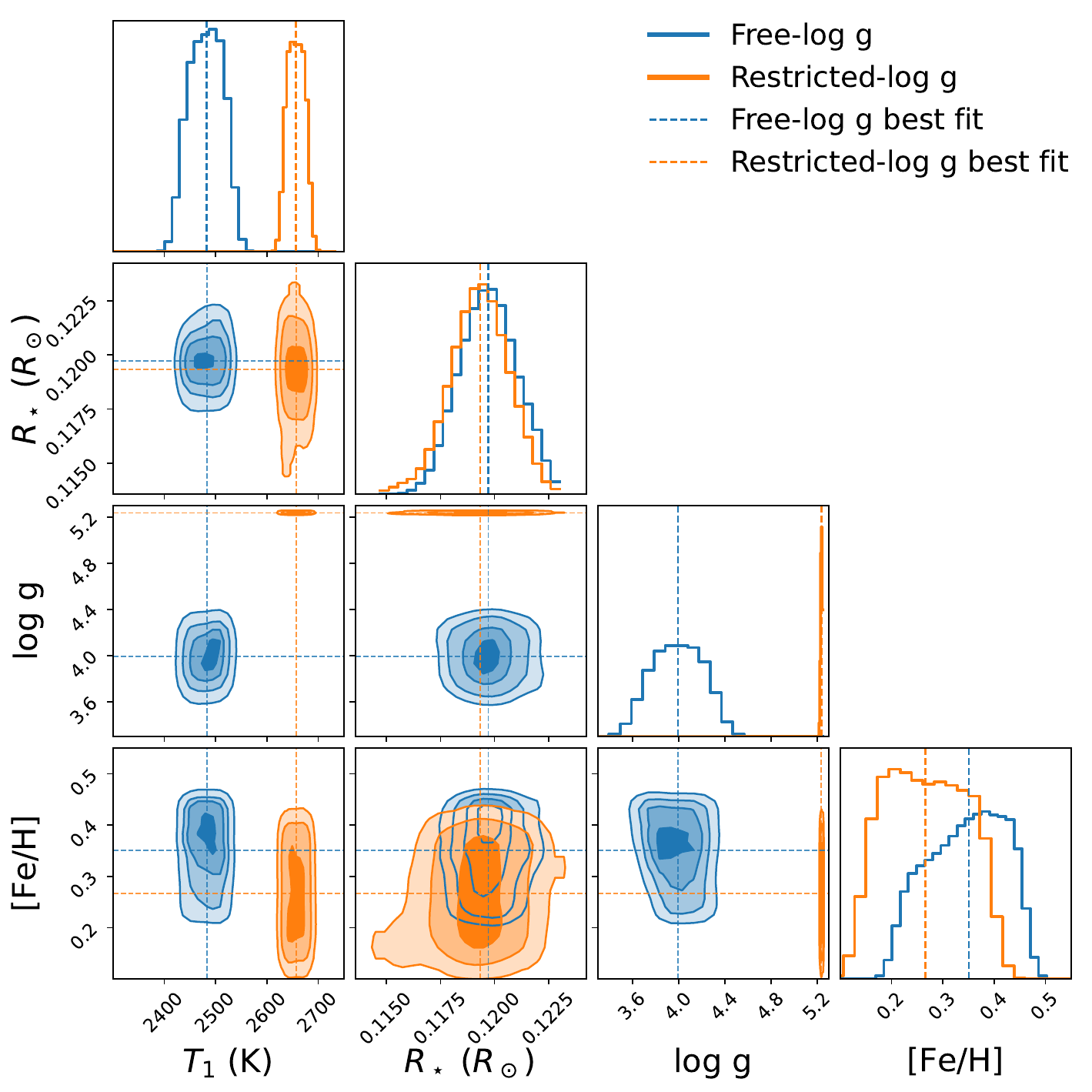}
    \caption{Example corner plots showing the posterior distributions of the retrieved stellar parameters for TRAPPIST-1 using the PHOENIX/NewEra model grid. 
    The blue and orange distributions correspond to the free-$\log g$ and restricted-$\log g$ scenarios, respectively.  
    The parameters shown here are the effective temperature ($T_1$), stellar radius ($R_\star$), surface gravity ($\log g$), and metallicity ([Fe/H]).
    The dashed lines in each one-dimensional posterior distribution mark the best-fit parameter values. The retrieval results for the full sample, including TRAPPIST-1, are summarized in Table~\ref{tab:retrieved_params_free_fixed_logg}.
    } 
    \label{fig:1-component retrieval Trappist-1 corner plot}
\end{figure}

As a representative example of the posterior distributions obtained in these retrievals, Figure~\ref{fig:1-component retrieval Trappist-1 corner plot} shows the corner plots for TRAPPIST-1 under the two scenarios.
Under the free-$\log g$ scenario, the retrieval yields a surface gravity of $\log g = 3.995^{+0.222}_{-0.217}$, substantially lower than the expected value for TRAPPIST-1 of $\log g \approx5.24$.
When $\log g$ is restricted to the expected high-gravity range, the posterior is constrained to $\log g = 5.238^{+0.002}_{-0.002}$, and the retrieved photospheric temperature increases to $T_1 = 2657^{+17.3}_{-17.0}$~K, compared with $T_1 = 2483^{+31.8}_{-34.4}$~K in the free-$\log g$ case.

Figure~\ref{fig:1-component retrieval Trappist-1} compares the corresponding single-component spectral models with the TRAPPIST-1 NIRISS/SOSS mean flare-free and out-of-transit spectrum from 1.3~$\mu$m to 2.8~$\mu$m.
The black points represent the observed spectrum, while the purple curves show the best-retrieved PHOENIX/NewEra model spectra.
The left and right panels correspond to the free-$\log g$ and restricted-$\log g$ scenarios, respectively.
The lower subpanel in each case shows the residuals between the observed spectrum and the best-fit model.
For TRAPPIST-1, the freely retrieved low-$\log g$ solution provides a substantially better fit, yielding a reduced chi-square of $\chi^{2}_{\nu}=1.019$, compared with $\chi^{2}_{\nu}=3.348$ for the restricted-$\log g$ scenario.


\subsubsection{{Retrievals of 22 Late-M Dwarfs from IRTF/SpeX}}

{
Figure~\ref{fig:one_component_retrieval_all_M_dwarfs} shows the one-component spectral modeling of six of the 22 IRTF/SpeX late-M dwarfs in our sample over the wavelength range from 1.3~$\mu$m to 2.4~$\mu$m. The modeling results for the remaining 16 IRTF/SpeX late-M dwarfs are shown in Figure~\ref{fig:appendix_spex_retrievals} in Appendix~\ref{subsec:additional_spex_retrievals}.
}
The figure follows the same format as Figure~\ref{fig:1-component retrieval Trappist-1} but for the IRTF/SpeX spectra.
We further compare the modeling results from the two scenarios using the Bayesian information criterion \citep[BIC;][]{Schwarz1978}. 
We define the BIC difference as $\Delta{\rm BIC}_{\rm free-R} = {\rm BIC}_{\rm free} - {\rm BIC}_{\rm R}$, where "R" here denotes the restricted-$\log g$ scenario. 
Therefore, $\Delta{\rm BIC}_{\rm free-R}<0$ indicates that the observed spectrum favors the free-$\log g$ scenario, whereas $\Delta{\rm BIC}_{\rm free-R}>0$ indicates that the restricted-$\log g$ scenario is favored. 
We also apply the same analysis demonstrated above for TRAPPIST-1 to all other stars in the sample by estimating the $\chi^{2}_{\nu}$ values for both scenarios.
The resulting $\chi^{2}_{\nu}$ values, together with $\Delta{\rm BIC}_{\rm free-R}$, are listed in Table~\ref{tab:retrieved_params_free_fixed_logg}.



{
We also examined whether the 2300~K lower boundary of the PHOENIX/NewEra grid affects the retrieved $T_{\rm eff}$ values. 
Even for the coolest stars in our sample, the inferred temperatures and their posterior distributions remain well above the lower grid boundary, indicating that the 2300~K limit does not truncate the temperature posteriors or significantly affect their derived uncertainties.
}

\subsubsection{{Comparison of Free-$\log g$ and Restricted-$\log g$ Retrievals Across the Sample}}

\begin{figure}
    \centering
    \includegraphics[width=\textwidth, height=0.55\textheight, keepaspectratio]{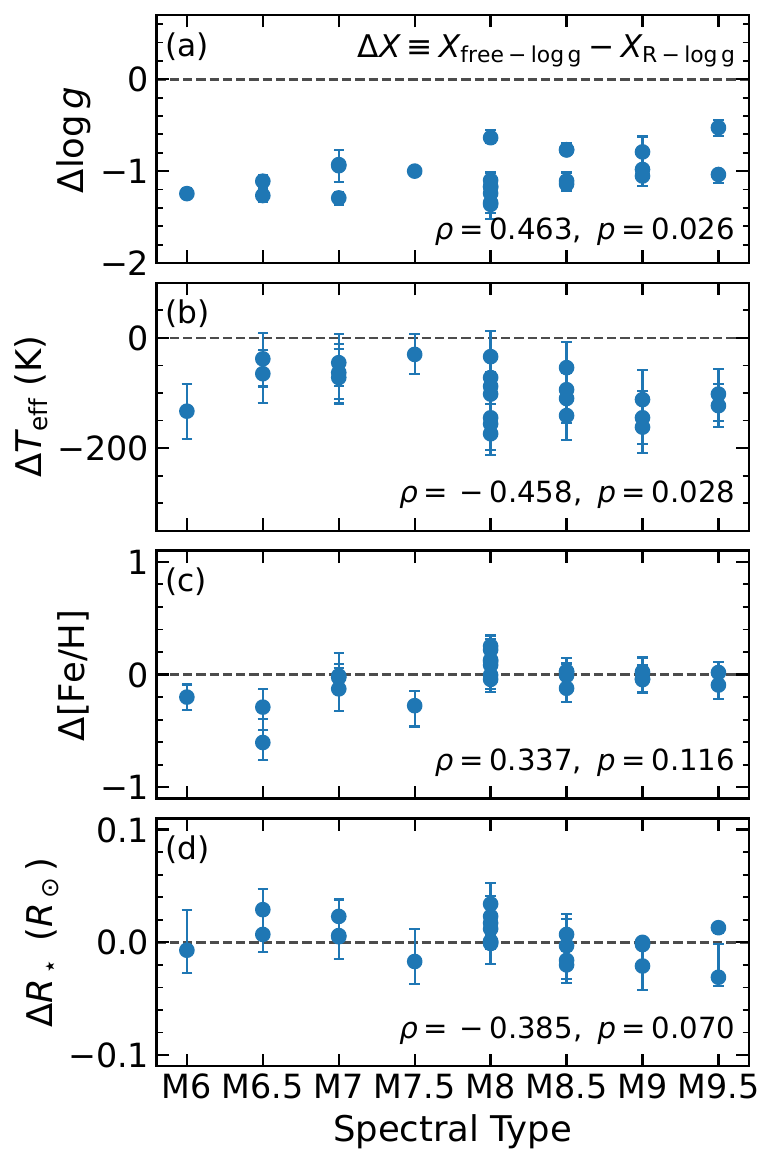}
    \caption{
    Differences in the retrieved stellar parameters between the free-$\log g$ and restricted-$\log g$ fitting scenarios as a function of spectral type, defined as $\Delta X = X_{\rm free-\log g} - X_{\rm R-\log g}$. From top to bottom, the panels show $\Delta \log g$, $\Delta T_{\rm eff}$, $\Delta{\rm [Fe/H]}$, and $\Delta R_\star$. The Spearman correlation coefficients and corresponding $p$-values are shown in each panel. 
    } 
    \label{fig:retrieved_parameters_between_freelogg_and_Rlogg}
\end{figure}

{
Our results show that the NIR spectra of all late-M dwarfs in our sample favor the free-$\log g$ scenario, with $\Delta{\rm BIC}_{\rm free-R}<0$ for every star. 
To further examine the differences in the retrieved stellar parameters between the two scenarios across the sample, Figure~\ref{fig:retrieved_parameters_between_freelogg_and_Rlogg} shows $\Delta X = X_{\rm free-\log g} - X_{\rm R-\log g}$ as a function of spectral type. The $\Delta \log g$ values are negative for all stars, showing that the free-$\log g$ retrieval systematically gives lower surface gravities than the restricted-$\log g$ retrieval. We find a moderate positive correlation between $\Delta \log g$ and spectral type ($\rho=0.463$, $p=0.026$), indicating that the difference in retrieved $\log g$ becomes smaller toward later spectral types. The retrieved $T_{\rm eff}$ values are also systematically lower in the free-$\log g$ scenario, with $\Delta T_{\rm eff}<0$ for all stars, and show a moderate negative correlation with spectral type ($\rho=-0.458$, $p=0.028$).}

{
This systematic shift likely reflects a covariance between $T_{\rm eff}$ and $\log g$ in the spectral modeling. When $\log g$ is restricted to the expected high-gravity range, the retrieval tends to compensate by favoring a higher $T_{\rm eff}$.
In contrast, the differences in [Fe/H] and stellar radius are distributed around zero, with no consistent tendency for either parameter to be higher or lower in the free-$\log g$ scenario compared with the restricted-$\log g$ scenario.
This consistent preference for the free-$\log g$ scenario suggests that pseudo-continuum spectral modeling with current state-of-the-art stellar atmosphere grids may tend to retrieve underestimated surface gravities for late-M dwarfs, with corresponding systematic shifts toward lower $T_{\rm eff}$ values.}

\subsubsection{{Comparability of the NIRISS/SOSS and IRTF/SpeX Retrievals}}

{
The NIRISS/SOSS and IRTF/SpeX spectra analyzed here have different data precisions, adopted wavelength coverages, and spectral resolving power~(see Section~\ref{sec:observation_and_data_reduction}). 
We discuss and examine 
the comparability of the retrieval results obtained from the two instruments.
}

{
Regarding the difference in data precision, we do not expect it to systematically bias the retrieved stellar parameters because the observational uncertainty of each spectral point is explicitly included in the likelihood. The lower precision of the SpeX spectra may therefore be reflected mainly in the fitted error-inflation term, which is generally larger for the SpeX spectra than for the NIRISS/SOSS spectrum.}

{
Although the SpeX spectra have a higher spectral resolving power, this difference is explicitly accounted for in our modeling by convolving the PHOENIX/NewEra spectra to match the spectral resolution of each observed dataset before fitting. 
Moreover, our spectral retrievals are primarily constrained by the broad near-infrared pseudo-continuum and molecular-band morphology rather than individual narrow spectral lines. 
We therefore do not expect the difference in spectral resolving power between NIRISS/SOSS and IRTF/SpeX to introduce a significant systematic bias in the retrieved stellar parameters.
}

{
To test the effect of the different wavelength coverage, we repeated the free-$\log g$ retrieval of TRAPPIST-1 after restricting the NIRISS/SOSS spectrum to the same wavelength coverage used for our SpeX M~dwarfs (see Section~\ref{subsec:wcoverage_test_for_retrievals} for more details).
As a result, the stellar parameters retrieved from the NIRISS/SOSS spectrum restricted to the SpeX-like wavelength coverage remain consistent within uncertainties with those obtained from the 1.3--2.8~$\mu$m retrieval adopted in our main analysis, as shown in Figure~\ref{fig:1_corner_plots_wcoverage_test}.  
This agreement suggests that the retrieval results obtained from the NIRISS/SOSS spectrum of TRAPPIST-1 and the IRTF/SpeX spectra of the other late-M dwarfs can be reasonably compared in the subsequent analyses in this study, despite their different wavelength coverage.
}




\subsection{Gravity-Sensitive Spectral Indices and Free-$\log g$ Retrievals}
\label{subsec:gravity_spi_vs_free_logg}


The gravity-sensitive spectral indices we measured for our late-M~dwarf sample are summarized in Table~\ref{tab:gravity_sensitive_indices}.
Figure~\ref{fig:free-logg-vs-gravity_spectral_index} compares the measured gravity-sensitive spectral indices with the surface gravities retrieved from the free-$\log g$ spectral modeling. 
The retrieved $\log g$ values show moderate positive correlations with the FeH$_z$, FeH$_J$, and K I$_J$ indices, with Pearson correlation coefficients of $r=0.576$, $0.576$, and $0.460$, respectively. 
The corresponding p-values are $0.004$, $0.004$, and $0.027$, indicating statistically significant correlations for these three indices. 

Because our sample size is modest, we also compute Spearman rank correlation coefficients, which test for monotonic rather than strictly linear relationships. 
The Spearman coefficients of these three indices are also positive, with $\rho=0.603$, $0.608$, and $0.474$, and corresponding p-values of $0.0023$, $0.0021$, and $0.0222$, respectively.  
The VO$_z$ index also shows a positive trend, with $r=0.455$ and $p=0.0381$, although the Spearman result is weaker ($\rho=0.391$, $p=0.0797$). 
The H-cont index shows a moderate negative correlation with retrieved $\log g$, with $r=-0.474$ and $p=0.0224$, and a consistent Spearman correlation of $\rho=-0.450$ with $p=0.0314$. 

The consistency between the Pearson and Spearman results for FeH$_z$, FeH$_J$, K\,\textsc{i}$_J$, and H-cont suggests that these correlations are not driven solely by the assumed linear form or by a small number of high-leverage points. 
The VO$_z$ index shows the same positive correlation direction in both tests, but its weaker Spearman significance indicates that this trend should be interpreted more cautiously.

We further fit an empirical linear relation between each gravity-sensitive spectral index and the retrieved or ``observed'' surface gravity,
\begin{equation}
\log g_{\rm obs} = a I + b,
\label{eq:logg-to-spectral_indices}
\end{equation}
where $I$ is the measured spectral index (e.g., FeH$_z$, FeH$_J$, K I$_J$, etc), and $a$ and $b$ are the fitted slope and intercept, respectively.
The best-fit relations and coefficients are also shown in Figure~\ref{fig:free-logg-vs-gravity_spectral_index}.
These empirical correlations can be applied to late-M dwarfs spanning spectral types M6--M9.5 to estimate the $\log g$ values preferred by the observed gravity-sensitive spectral features.

These trends suggest that the freely retrieved surface gravities are connected to the gravity-sensitive features present in the observed NIR spectra.
The correlations between the independently measured gravity-sensitive spectral indices and the underestimated $\log g$ values further support that the low surface gravities preferred by the pseudo-continuum fitting using the PHOENIX/NewEra model grid may not be purely numerical artifacts, but instead reflect genuine properties of the observed NIR spectra of the late-M dwarfs.


\subsection{Connection Between Underestimated Surface Gravity and Magnetic Activity}



The primary objective of this study is to investigate the connection between the discrepancies in the retrieved $\log g$ values and magnetic activity.
We find strong anti-correlations between the freely retrieved $\log g$ values and the two magnetic activity indicators analyzed in this work, flare luminosity ($L_{\rm flare, TESS}$) and X-ray luminosity ($L_X$), for the late-M dwarfs in our sample (see Figure~\ref{fig:flare_and_X-ray-vs-retrieved_logg}).
The general trend is that stars with higher activity levels, indicated by higher $L_X$ or $L_{\rm flare, TESS}$, tend to exhibit NIR spectral features that are reproduced by lower-$\log g$ models.
These anti-correlations are statistically significant, with Pearson correlation coefficients of $r=-0.931$ and $-0.948$ for the $\log g$--$\log L_{\rm flare, TESS}$ and $\log g$--$\log L_X$ relationships, respectively. The corresponding p-values are $3.78\times10^{-6}$ and $1.0\times10^{-4}$. We also compute Spearman rank correlation coefficients to test whether these trends remain significant without assuming a strictly linear relationship. 
The Spearman coefficients are $\rho=-0.945$ and $-0.950$, also indicating statistically significant relationships with p-values of $1.12\times10^{-6}$ and $8.76\times10^{-5}$ for the $\log g$--$\log L_{\rm flare, TESS}$ and $\log g$--$\log L_X$ relationships, respectively.

We derive the empirical linear relations between the freely retrieved $\log g$ values and the two magnetic activity indicators in logarithmic space.
For the flare luminosity, the resulting relation is
\begin{equation}
\log g = (8.324\pm0.561) - (0.165\pm0.023)\log_{}(L_{\rm flare, TESS}).
\label{eq: logg-vs-flare_lum}
\end{equation}
For the X-ray luminosity, we obtain
\begin{equation}
\log g = (10.458\pm0.706) - (0.243\pm0.028)\log_{}(L_X).
\label{eq:logg-vs-X_lum}
\end{equation}
We note that the $\log g$--$\log (L_X)$ relation was derived by treating the reported upper limits on $L_X$ as the adopted $L_X$ values for the late-M dwarfs without X-ray detections.
{
As a robustness check, we repeated the correlation analysis using only the six stars with X-ray detections. 
The strong anti-correlation persists, with Pearson $r=-0.958$ ($p=0.0026$) and Spearman $\rho=-0.899$ ($p=0.015$), indicating that the observed trend is not driven by the treatment of the upper limits.
}

{
We also note that the stellar spectra and activity measurements are not contemporaneous.
These correlations therefore trace relationships between the characteristic activity levels and spectral properties of the stars across the sample, rather than short-term spectral responses to individual flare events. 
}

\section{Discussion} \label{sec:discussion}

Our results show that the near-infrared spectra of the late-M~dwarfs in our sample are better reproduced by PHOENIX/NewEra models with unrealistically low surface gravities. 
This suggests that the retrieved $\log g$ values may compensate for atmospheric physics or spectral-model uncertainties that are not fully captured by the current models. 
In particular, uncertainties in molecular-line pressure broadening may affect the strengths of the broad molecular features and contribute to the preference for low-$\log g$ solutions. 
At the same time, the strong anti-correlations between the retrieved $\log g$ values and both flare and X-ray luminosities suggest that magnetic activity may also contribute to the observed low-gravity-like spectral morphology.


In this section, we first discuss whether uncertainties in pressure broadening may contribute to the correlations between the retrieved $\log g$ values and the gravity-sensitive spectral indices shown in Figure~\ref{fig:free-logg-vs-gravity_spectral_index}. 
We next consider magnetic heating as a possible explanation for the lower retrieved $\log g$ values found for more active stars.  
We first use the available NIRISS/SOSS spectra of TRAPPIST-1 to examine whether individual energetic flares produce detectable short-timescale changes in magnetically sensitive photospheric features. 
We then present a simplified radiative-transfer experiment in which we construct a series of heated T--P profiles representing different activity-related heating scenarios and examine how the modeled spectra and gravity-sensitive indices respond to the different heating scenarios.

\subsection{Pressure Broadening and the Low-$\log g$ Solutions}
\label{sec:pressure_broadening}

One possible explanation for the unrealistically low retrieved $\log g$ values is the treatment of pressure broadening in current stellar atmosphere models. 
Calculations with MPS-ATLAS show that surface gravity has only a small effect on the temperature structure as a function of Rosseland optical depth, except in regions affected by convection and overshooting \citep{2022A&A...666A..60K}. 
However, surface gravity still changes the gas pressure and density at a given Rosseland optical depth. 
Lower-gravity models therefore have lower gas pressure and density and weaker collisional broadening, including van der Waals broadening.

Late-M~dwarf spectra contain many saturated molecular bands and pressure-sensitive atomic lines. 
Stronger pressure broadening increases the effective widths of these features and can make the broad absorption bands stronger. 
If the adopted pressure broadening is overestimated, the modeled FeH, VO, and K~I features may be too strong and may also respond too weakly to changes in $T_{\rm eff}$ and $[\mathrm{Fe/H}]$. 
For a star with weaker observed absorption features, varying $T_{\rm eff}$ or $[\mathrm{Fe/H}]$ may therefore not be enough to reproduce the spectrum. 
The retrieval may instead prefer a lower $\log g$, which reduces the atmospheric pressure and weakens the pressure broadening.
This behavior could help explain the correlations shown in Figure~\ref{fig:free-logg-vs-gravity_spectral_index}.

{
\citet{2026ApJ...997L...8G} showed that reducing the adopted van der Waals broadening improves the agreement between synthetic spectra and the observed FeH Wing--Ford band and near-infrared pseudo-continuum of TRAPPIST-1, suggesting that the adopted pressure broadening may be overestimated in current models.
Because collisional broadening depends on gas pressure, magnetic pressure could also weaken these features by reducing the gas pressure within magnetically active regions.
}

{
To estimate whether this effect is large enough, we consider the FeH Wing--Ford band near $0.998~\mu$m as an example. 
This feature forms relatively deep in M-dwarf atmospheres, around $\log\tau\sim-1$ \citep{2019A&A...629A..83A}. 
In a representative PHOENIX/NewEra atmosphere with $T_{\rm eff}~=~2500$~K and $\log g~=~5$, this corresponds to a gas pressure of approximately 1~bar. 
By comparison, our free-$\log g$ fit for TRAPPIST-1 favors $\log g\sim4$, for which the gas pressure at a comparable optical depth is of order 0.1~bar.
For a magnetic flux tube in horizontal pressure balance with its surroundings \citep{2026ApJ..1008...24S}, $P_{\rm gas,out}=P_{\rm gas,in}+P_B$, where $P_{\rm gas,out}$ and $P_{\rm gas,in}$ are the gas pressures outside and inside the magnetic flux tube, respectively, $P_B=B^2/(8\pi)$ is the magnetic pressure, and $B$ is magnetic field strength.
Reducing the gas pressure from $\sim1$~bar to $\sim0.1$~bar requires a local magnetic field of $B\sim4$--$5$~kG.
}

{
However, reproducing the globally low-gravity-like spectrum would require such strong fields to cover a large fraction of the stellar surface, implying a surface-averaged magnetic flux of order several kG. 
This is much larger than the measured $Bf\sim600$~G for TRAPPIST-1 \citep{2010ApJ...710..924R}.
Therefore, magnetic pressure alone is unlikely to reduce the gas pressure, and thus the pressure broadening, enough to reproduce the observed low-gravity-like spectral morphology of TRAPPIST-1.}

{
Given the strong correlations between the freely retrieved $\log g$ values and both flare and X-ray luminosities, we consider that magnetic activity may alter the atmospheric structure more substantially than through magnetic pressure alone, possibly through heating associated with flare activity. 
In the following sections, we explore whether such changes in the atmospheric thermal structure can contribute to the observed low-gravity-like spectral morphology.
}

\subsection{Weak Variability in Magnetically Sensitive Features during Large Flares}
\label{subsec::weak_flare_phase_variability_in_FeH_and_Ti8380}

Here, we consider magnetic heating as a possible contributor to the observed activity-dependent low-$\log g$ behavior.
In this context, downward-directed energetic particles and radiation during flare events may deposit energy in the lower atmosphere and modify its thermal structure.
Longer-lasting heating may also be maintained by widespread active regions or frequent lower-energetic flares. 
These activity-related changes in the atmospheric T--P structure could cause the emergent near-infrared spectra of late-M~dwarfs to resemble those of lower-gravity model atmospheres, even when their true surface gravities remain high.
Under this hypothesis, an important question is whether the altered atmospheric structure is produced by transient heating during individual energetic flares, or whether it reflects a more persistent magnetic-heating process, such as widespread active regions or frequent low-level flare activity across the stellar atmosphere.  
We investigate this question using the available NIRISS/SOSS spectral observations of TRAPPIST-1.

Six flares were detected on TRAPPIST-1 during its nine NIRISS/SOSS visits from GO~2589 and GTO~1201, obtained between 2022 and 2023 \citep{2023ApJ...959...64H}.
To examine whether individual energetic flares produce detectable changes in magnetically sensitive spectral features, we focus on the FeH$_z$ index.
The FeH Wing--Ford band near $0.998~\mu{\rm m}$ is known to be sensitive to photospheric magnetic fields in late-M~dwarfs through Zeeman-induced changes in its spectral morphology \citep[e.g.,][]{2007ApJ...656.1121R}, and this feature forms relatively deep in the photosphere of M~dwarfs \citep{2019A&A...629A..83A}.
Therefore, the time-resolved FeH$_z$ index provides a useful test of whether individual energetic flares produce detectable changes in this magnetically relevant photospheric spectral region.

For each visit in which a flare was detected, we measure the FeH$_z$ index for every spectrum in the time series using the same method described in Section~\ref{subsec: measure spectral gravity sensitive indices}.
This produces a time-resolved ${\rm FeH}_z$ time series for each flare-detected visit, which we compare with the corresponding white-light and H$\alpha$ flare light curves to test whether the FeH$_z$ feature shows detectable variation during individual flare events.

We do not observe a clear variation in the FeH$_z$ index during these flare events.
Figure~\ref{fig:trappist1_niriss_soss_flares_and_FeH} shows representative examples of the time series of the white-light curve, H$\alpha$ light curve, and FeH$_z$ index during the two most energetic flares observed among the TRAPPIST-1 NIRISS/SOSS visits. 
One flare occurred during the GTO 1201 visit on 2023~July~22, and the other occurred during the visit on 2023~October~31. 
Both flares show clear brightness enhancements in the white-light and H$\alpha$ light curves, while no clear variation is observed in the FeH$_z$ index during the flare events with the durations of 1.0 to 1.6 hours.  
We also find no clear pre-flare variation in the FeH$_z$ index within about 0.5--1 hour before the flares, nor any delayed response within about 1.5--2.1 hours after the flares.
For the other smaller flares, we do not see clear variations in FeH$_z$ either.

The lack of clear variation in the FeH$_z$ index during the flare events may indicate that the transient flare energy deposition does not strongly affect the deep photospheric layers traced by this feature or that the response of FeH$_z$ occurs on a longer timescale not covered by the observations. 
To further test whether magnetically sensitive features formed at different photospheric depths respond to the flares, we also examine the Ti\,\textsc{i}~8380 index. 
The Ti\,\textsc{i} lines near 8380~\AA\ are magnetically sensitive atomic features and are expected to form higher in the photosphere than the FeH Wing--Ford band \citep{2019A&A...629A..83A, 2021A&ARv..29....1K}.

We estimate the Ti\,\textsc{i}~8380 index using Equation~\ref{eq:spectral_index}, with the adopted parameters $\lambda_{\rm line}=0.8380~\mu{\rm m}$, $\lambda_{\rm cont1}=0.820~\mu{\rm m}$, and $\lambda_{\rm cont2}=0.856~\mu{\rm m}$. We apply a bandwidth of $0.013~\mu{\rm m}$ to the line and both continuum regions.
The time series of the Ti\,\textsc{i}~8380 index during the two most energetic flares observed on TRAPPIST-1 with NIRISS/SOSS are also shown in Figure~\ref{fig:trappist1_niriss_soss_flares_and_FeH}. 
Similar to FeH$_z$, the Ti\,\textsc{i}~8380 index does not show a clear variation during the flare events. 
This suggests that these flares do not produce a detectable short-timescale change in the magnetically sensitive photospheric features examined here, including FeH$_z$ formed in deeper photospheric layers and Ti\,\textsc{i}~8380 formed in higher photospheric layers.

Previous studies have suggested that the cumulative effect of small-scale magnetic activity can contribute significantly to the apparently quiescent emission of active stars, especially when low-energy events occur frequently enough to dominate the time-averaged energy budget \citep[e.g.,][]{1991SoPh..133..357H,2002ApJ...580.1118K,2003A&A...403..155G,2015csss...18..130T}.
In this picture, the relevant quantity may not be only the energy of an individual magnetic event, but also the effective surface coverage, or filling factor, of magnetically heated regions.
In this context, our observations of TRAPPIST-1 suggest that the main source of magnetic heating may not be the detectable, relatively energetic flares, but instead smaller and more frequent events that cover a larger time-averaged filling factor and continuously heat the stellar atmosphere.
This interpretation is consistent with the results of \citet{Berardo2026}, who found that HST/STIS Ly$\alpha$ and VLT $g'$ observations of TRAPPIST-1 point to frequent ${\sim}10^{29}$\,erg flares on subhour timescales as an important contributor to the star's persistent activity.

\subsection{A Simplified Model to Test Magnetic Heating}
\label{subsec: A toy moodel to test magnetic heating}
A possible way to test the magnetic-heating hypothesis is to generate a photospheric heating spectral grid based on the stellar parameters of the target star.
For example, for TRAPPIST-1, one can start from a reference atmosphere with stellar parameters close to the expected values of the star, such as $T_{\rm eff}\approx2500~{\rm K}$, $\log g \approx 5.0$--$5.2$, and approximately solar metallicity.
The reference atmosphere can be taken from existing stellar atmosphere grids such as PHOENIX/NewEra or PHOENIX ACES, which provide synthetic spectra and atmosphere structures over a broad range of stellar parameters \citep[e.g.,][]{2013A&A...553A...6H}.
The original temperature-pressure structure of this atmosphere can then be modified by depositing different levels of magnetic heating at selected atmospheric layers, especially in the upper photosphere or lower-pressure regions.
After the thermal structure is modified, radiative-transfer calculations can be carried out, for example with stellar spectral-synthesis frameworks such as \texttt{MPS-ATLAS} \citep[e.g.,][]{2021A&A...653A..65W} or \texttt{petitRADTRANS} \citep[e.g.,][]{Molliere2019, 2024JOSS....9.7028B}, to generate a series of synthetic spectra corresponding to different heating strengths and heating depths. From these modeled spectra, we examine whether pressure- or gravity-sensitive spectral features vary with the adopted heating strength and heating depth.

Developing such a physically robust heated-atmosphere spectral grid requires substantial modeling effort and is beyond the scope of this paper. 
Instead, we perform a simplified proof-of-concept test to illustrate how the spectral features discussed in this work respond to modified atmospheric temperature structures. 
Synthetic spectra are then generated from these modified structures to examine whether gravity-sensitive features, including FeH$_z$, FeH$_J$, K\,\textsc{i}$_J$, VO$_z$, and the H-band continuum index, vary systematically with the imposed heating.


Starting from the original PHOENIX ACES T--P profile with $T_{\rm eff}=2500~{\rm K}$, $\log g=5.0$, and $[\mathrm{Fe/H}]=0$, we construct a set of toy heated profiles by adding a pressure-dependent temperature perturbation,
\begin{equation}
T_{\rm heated}(P) = T_{\rm quiet}(P) + \Delta T(P),
\end{equation}
where
\begin{equation}
\Delta T(P) =
\Delta T_{\rm deep} +
\frac{\Delta T_{\rm upper}}
{1+\left(P/P_{\rm trans}\right)^{\alpha}} .
\end{equation}
Here, $\Delta T_{\rm upper}$ controls the heating amplitude in the low-pressure upper atmosphere, $\Delta T_{\rm deep}$ represents the imposed temperature offset that remains at higher pressures, $P_{\rm trans}$ sets the pressure level where the perturbation transitions between the upper and deeper atmosphere, and $\alpha$ controls the sharpness of this transition. 
We adopt $\alpha=1.5$ and generate 10 illustrative heating scenarios with $\Delta T_{\rm upper}=100$--$1000~{\rm K}$, $\Delta T_{\rm deep}=20$--$200~{\rm K}$, and $P_{\rm trans}=10^{-3}$--$10^{-0.5}~{\rm bar}$, ordered from weak to strong heating cases.
These perturbation amplitudes are broadly motivated by radiative--hydrodynamic models of M-dwarf flares from the literature, which show that chromospheric heating and radiative backwarming can modify the lower-atmosphere thermal structure; for example, the F11 model discussed by \citet{2015SoPh..290.3487K} produces a photospheric temperature increase of order $\sim10^3~{\rm K}$.
At the same time, the detailed mapping between flare energy deposition and the resulting layer-by-layer perturbation of the atmospheric T--P structure remains uncertain. 
Our imposed perturbations are therefore intended to provide a simple, weaker, spatially and/or temporally averaged representation of heating associated with unresolved small-scale magnetic activity or microflares, rather than an instantaneous flare kernel. 
The adopted heating scenarios should therefore be interpreted as a sensitivity test of how activity-related changes in the atmospheric T--P structure affect the emergent spectral morphology, rather than as a self-consistent flare-atmosphere calculation.
Our modified, heated T--P profiles are shown in panel (a) of Figure~\ref{fig:simple_modified_T-P_magnetic_heat_atmosphere_spectra}. 
The heating scenarios are ordered from 1 to 10, where Scenario~1 represents the weakest imposed heating and Scenario~10 represents the strongest imposed heating.

For each heated T--P profile, we compute an illustrative emergent spectrum at $R=1000$ over $0.9$--$2.8~\mu{\rm m}$ using a simplified radiative-transfer setup with \texttt{petitRADTRANS}.
The heated profiles are first interpolated onto a pressure grid spanning $10^{-6}$--$10^{2}~{\rm bar}$. 
We then adopt a fixed, vertically uniform composition for all heating scenarios, so that any spectral differences arise only from the imposed changes in the atmospheric temperature structure. 
The background atmosphere is dominated by H$_2$ and He. The included opacity sources are H$_2$O, CO, CH$_4$, NH$_3$, TiO, VO, FeH, Na, K, and additional neutral atomic metal lines from Ti, Fe, Si, Al, and Mg, together with H$^-$ continuum-related species, atomic H, and free electrons. Collision-induced absorption from H$_2$--H$_2$ and H$_2$--He is also included where applicable. 
The adopted volume mixing ratios are converted to mass fractions for the radiative-transfer calculation using the corresponding molecular or atomic weights.
We note that the adopted volume mixing ratios are used only for this toy radiative-transfer experiment. 
They were empirically chosen because, when applied to the quiet PHOENIX ACES T–-P atmospheric structure, they provide a reasonable match to the corresponding PHOENIX/NewEra spectrum over the wavelength range of 0.9--2.8~$\mu \rm m$ (Figure~\ref{fig:simple_RT_spectrum_of_quiet_TP}). 
These values are therefore intended as an effective opacity prescription, and are not meant to represent a chemically self-consistent abundance structure for late-M dwarfs.
The full list of adopted volume mixing ratios is provided in Table~\ref{tab:toy_vmr}.

The resulting modeled spectra are shown in panel (b) of Figure~\ref{fig:simple_modified_T-P_magnetic_heat_atmosphere_spectra}. The most obvious change is seen in the broad H$_2$O absorption features at about $1.3$--$1.5~\mu{\rm m}$, $1.75$--$2.05~\mu{\rm m}$, and $2.3$--$2.8~\mu{\rm m}$, which become shallower for the stronger heating scenarios.
The FeH$_z$ feature near $0.998~\mu{\rm m}$ also shows a noticeable change. 
Since the surface gravity is fixed in this experiment, this variation is not produced by changing $\log g$, but results from the imposed heating of the atmospheric structure. As shown in panel (c) of Figure~\ref{fig:simple_modified_T-P_magnetic_heat_atmosphere_spectra}, the FeH$_z$ feature becomes shallower for the stronger heating scenarios.

To quantify these changes, we estimate the spectral indices of FeH$_z$ and four other gravity-sensitive features discussed in this study (i.e., VO$_z$, FeH$_J$, K\,\textsc{i}$_J$, and H-cont), from the modeled spectra using the same method described in Section~\ref{subsec: measure spectral gravity sensitive indices}.
As a result, the FeH$_z$ index changes from $1.568\pm0.133$ in the weakest heating scenario to $1.123\pm0.024$ in the strongest heating scenario. 
Systematic changes between the weakest and strongest heating scenarios are also seen in the other gravity-sensitive features: from $1.321\pm0.088$ to $1.114\pm0.034$ for FeH$_J$, from $1.165\pm0.043$ to $1.088\pm0.020$ for K\,\textsc{i}$_J$, from $1.002\pm0.007$ to $1.019\pm0.005$ for H-cont, and from $1.072\pm0.059$ to $0.998\pm0.021$ for VO$_z$ (see Figure~\ref{fig:gravity_sensitive_indices_of_different_heated_spectra}).

These behaviors are broadly consistent with the correlations found in our retrieval results between the free-$\log g$ values, the observed gravity-sensitive spectral indices, and the magnetic activity indicators.
Since all modeled spectra are computed at the same input surface gravity of $\log g=5.0$, this toy experiment shows that changes in the atmospheric thermal structure can modify gravity-sensitive spectral indices without requiring a true change in $\log g$. 
Therefore, the underestimated surface gravities retrieved for active late-M dwarfs may not reflect intrinsically low surface gravity. 
Instead, they may indicate that their atmospheres are dynamically and thermally modified by persistent magnetic activity, such as microflare heating (based on the in-flare spectral analysis discussed in Section~\ref{subsec::weak_flare_phase_variability_in_FeH_and_Ti8380}), causing emergent spectra to mimic lower-gravity spectral behavior.

\subsection{Implications for Earlier M Dwarfs}
\label{subsec:earlier_m_dwarfs}


{
It is interesting to consider whether the activity-dependent low-gravity-like spectral behavior found here extends to earlier M dwarfs, for which similarly underestimated surface gravities from spectral modeling have also been reported.
For example, \citet{2022MNRAS.516.3802C} fitted high-resolution near-infrared SPIRou spectra of 44 early-mid M~dwarfs with synthetic spectra based on MARCS atmosphere models \citep{2008A&A...486..951G}.
They found a root-mean-square (RMS) difference of about 0.2\,dex between spectroscopic $\log g$ values and those inferred from stellar masses and radii, with the largest discrepancies occurring for the coolest stars.
For their six coolest targets, with $T_{\rm eff}\simeq2960$--3200\,K, the spectroscopic $\log g$ values were lower than the corresponding mass--radius estimates by approximately 0.26--0.66\,dex. 
Excluding these stars reduced the RMS difference to 0.11\,dex, leading \citet{2022MNRAS.516.3802C} to suggest that spectral fitting may underestimate $\log g$ toward the coolest end of their sample. 
However, they did not investigate whether these discrepancies are related to magnetic activity.}

{
Several stars in the \citet{2022MNRAS.516.3802C} sample have independently been reported to show magnetic activity.
GJ~1151 (M4.5V), for example, has detected coronal X-ray emission and an X-ray flare \citep{2020MNRAS.497.1015F}. 
Gl~905 (M5.0V), GJ~1289 (M4.5V), GJ~1151 (M4.5V), and GJ~1286 (M5.5V) have measured large-scale magnetic fields and temporal magnetic-field variations \citep{2024MNRAS.527.4330L}. 
In addition, optical flares have been detected in all four of these stars in TESS observations by \citet{2024AJ....168..234L}. 
These activity detections suggest that magnetic activity may also contribute to the underestimated $\log g$ values found for earlier M dwarfs, although this possibility remains to be tested.}

{Moreover, the empirical relations between the freely retrieved $\log g$ values and gravity-sensitive spectral indices derived in our work are specific to our M6--M9.5 sample. 
The gravity-sensitive spectral indices adopted in our work were originally defined and tested for ultracool dwarfs, primarily over the late-M and L-dwarf regime \citep{2013ApJ...772...79A}.
Therefore, these empirical relations derived from our late-M sample should not be directly extrapolated to earlier M dwarfs.}

{
In summary, previous work indicates that spectroscopic $\log g$ underestimation is not unique to late-M dwarfs, but there is not yet evidence that the same connection between low-gravity-like spectral morphology, gravity-sensitive spectral indices, and magnetic activity explored here for late-M dwarfs extends to earlier M dwarfs.
Applying a similar spectral-modeling and magnetic-activity analysis to earlier M dwarfs will help determine whether the activity-dependent low-gravity-like behavior identified here is also present across the broader M-dwarf sequence.
}


\section{Conclusions} \label{sec:conclusions}






In this work, we investigated the connection between underestimated surface gravities retrieved from models of near-infrared spectra and magnetic activity in late-M dwarfs. 
We analyzed a sample of 23 late-M dwarfs: TRAPPIST-1 observed with JWST NIRISS/SOSS, and 22 additional late-M dwarfs observed with IRTF/SpeX. 
We performed single-component spectral modeling using the PHOENIX/NewEra model grid and compared two fitting scenarios: one in which $\log g$ was allowed to vary freely, and one in which $\log g$ was restricted to the expected high-gravity range.
We also measured several gravity-sensitive spectral indices from the observed spectra, including FeH$_z$, VO$_z$, FeH$_J$, K\,\textsc{i}$_J$, and H-cont, and investigated their relationships with the retrieved $\log g$ values and magnetic activity indicators.
To examine the connection with magnetic activity, we compiled X-ray luminosities from the literature and derived flare luminosities from K2 and TESS observations for our sample late-M dwarfs as magnetic activity indicators.

The main findings of this work are summarized as follows.

\begin{figure*}
    \centering
    \includegraphics[width=\textwidth, height=0.4\textheight, keepaspectratio]{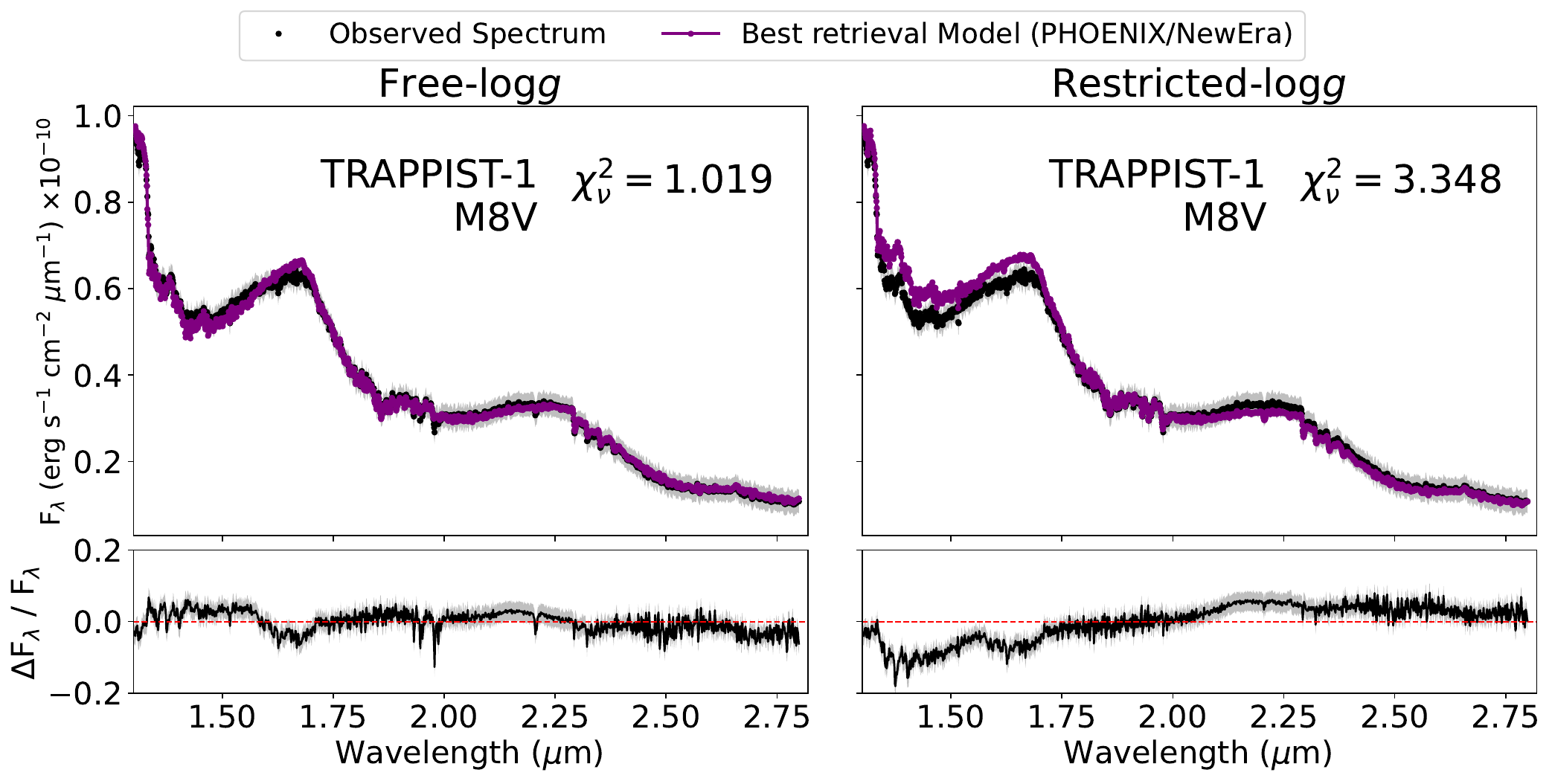}
    \caption{One-component spectral modeling of TRAPPIST-1's NIRISS/SOSS mean spectrum from 1.3~$\mu$m to 2.8~$\mu$m. 
    The black spectrum represents the observed spectrum of TRAPPIST-1. 
    The purple spectrum in the left panel shows the best-retrieved model for the free-$\log g$ scenario, while the purple spectrum in the right panel shows the best-retrieved model for the restricted-$\log g$ scenario.  
    {The gray shaded regions in both panels represent the $1\sigma$ uncertainties after including the fitted error-inflation term in addition to the photon-noise uncertainty.}
    In each panel, the lower subpanel shows the deviation between the observed spectrum and the best-retrieved model. 
    The best-retrieved parameters for both scenarios, the goodness of fit measured by the reduced chi-square ($\chi^{2}_{\nu}$), and the model comparison metric $\Delta \mathrm{BIC}_\mathrm{free-R}$ (BIC: Bayesian Information Criterion) are listed in Table~\ref{tab:retrieved_params_free_fixed_logg}.
    The best-retrieved model from the free-$\log g$ scenario is favored over that from the restricted-$\log g$ scenario, yielding a lower reduced chi-square of $\chi^{2}_{\nu}=1.019$ compared to $\chi^{2}_{\nu}=3.348$. 
    This is further supported by $\Delta{\rm BIC}_{\rm free-R}=-3576$.}
    \label{fig:1-component retrieval Trappist-1}
\end{figure*}
\begin{figure*}
    \centering
    \includegraphics[width=\textwidth, height=0.55\textheight, keepaspectratio]{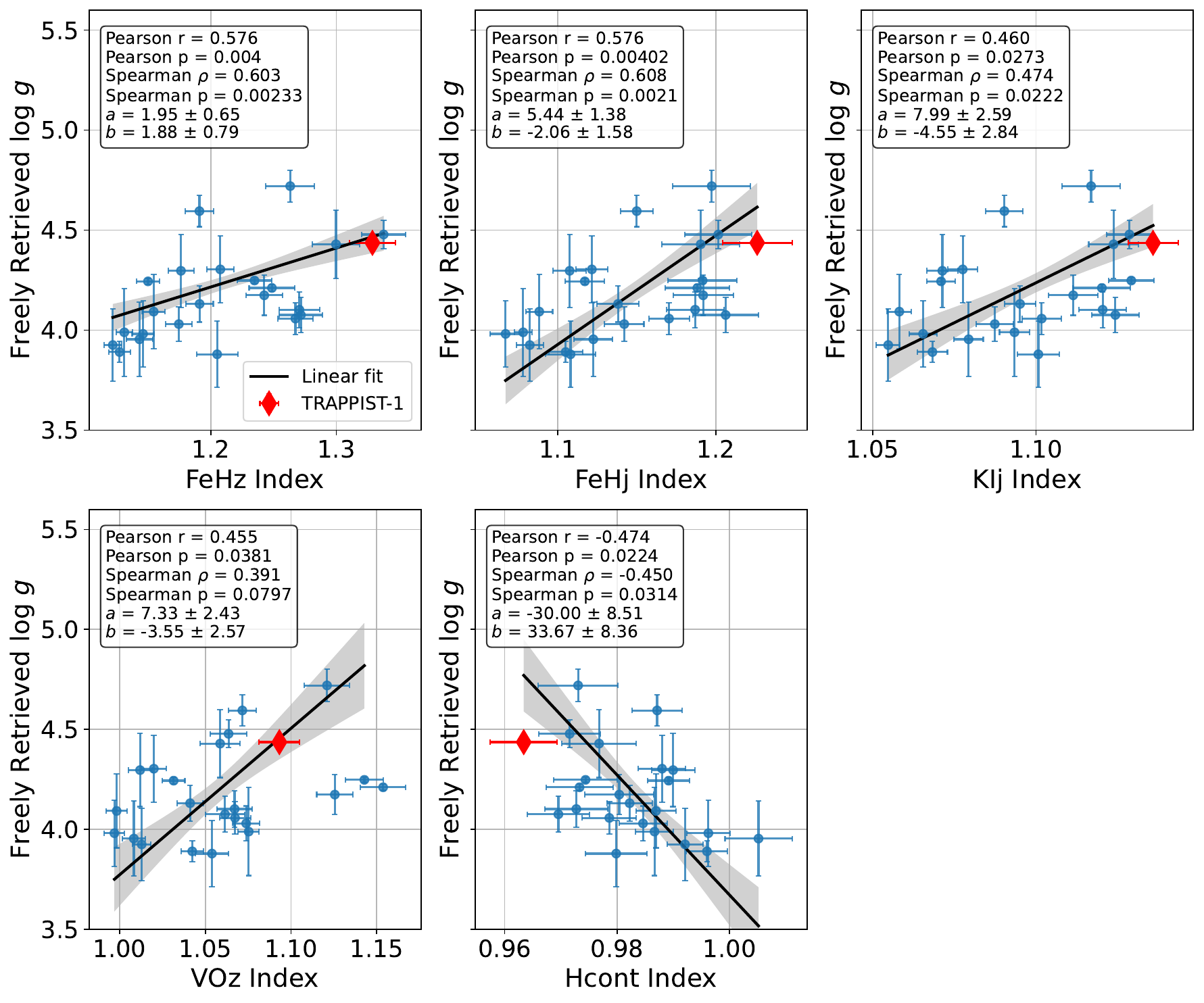}
    \caption{
    Correlations between the surface gravity $\log g$ retrieved from the free-$\log g$ spectral modeling scenario and the gravity-sensitive spectral indices of our late-M~dwarf sample. 
    TRAPPIST-1 is highlighted with a red diamond. 
    The FeH$_z$, FeH$_J$, and K I$_J$ indices show moderate positive correlations with the retrieved $\log g$. 
    The VO$_z$ index also shows a positive trend, 
    although the Spearman p-value ($p=0.0797$) is less significant than the Pearson p-value ($p=0.0381$). 
    The H-cont index shows a moderate negative correlation. 
    {
    The black solid line in each panel shows the empirical linear fit between the corresponding gravity-sensitive spectral index and the retrieved $\log g$ values, with the shaded region indicating the $1\sigma$ uncertainty of the fit. The fitted coefficients $a$ and $b$, following Equation~\ref{eq:logg-to-spectral_indices}, are labeled in each panel.
    }
    }
    \label{fig:free-logg-vs-gravity_spectral_index}
\end{figure*}
\begin{figure*}
    \centering
    \includegraphics[width=\textwidth, height=0.4\textheight, keepaspectratio]{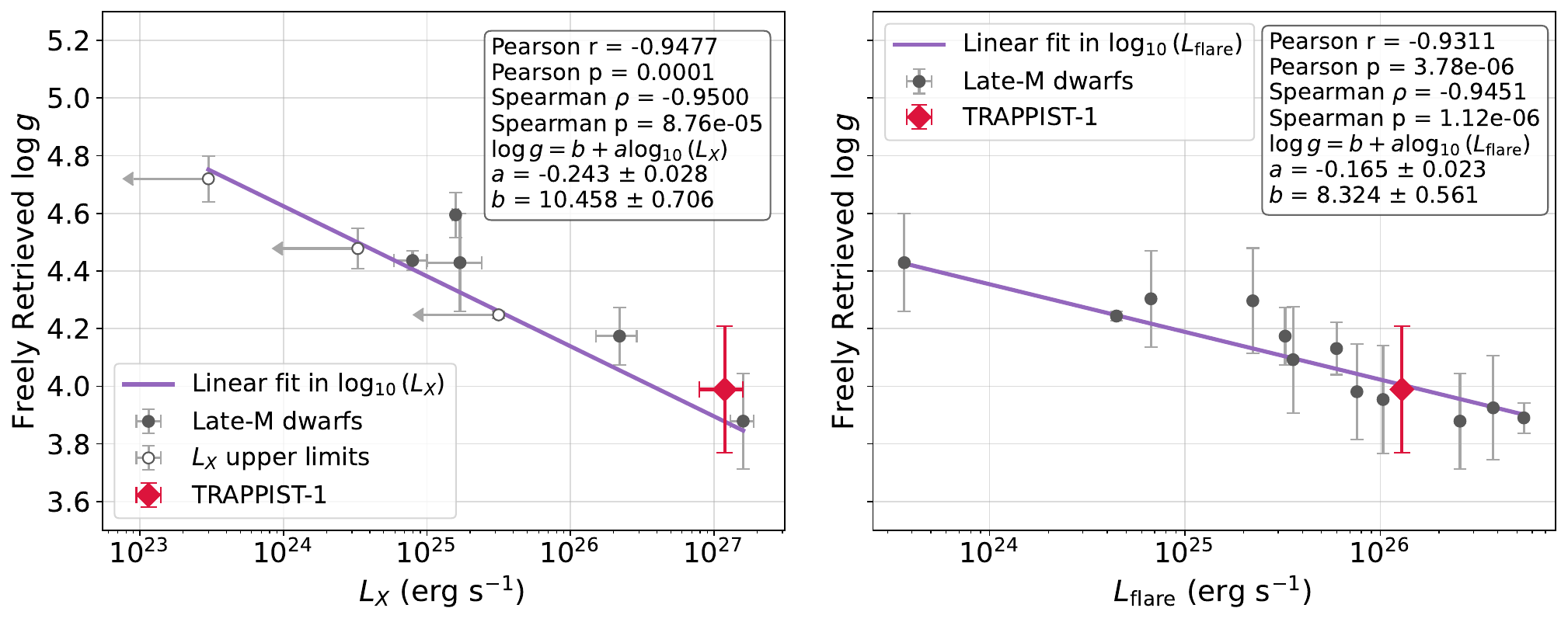}
    \caption{
    One-component retrieved $\log g$ for late-M dwarfs as a function of X-ray luminosity (left) and TESS-band flare luminosity (right).
    Gray circles show the comparison sample of late-M dwarfs, open symbols with arrows indicate $L_X$ upper limits, and the red diamond marks TRAPPIST-1.
    Purple lines show linear fits in log-luminosity space; 
    the corresponding Pearson correlation coefficients, Spearman rank coefficients, corresponding p-values, and best-fit parameters are listed in each panel.
    For both activity metrics, more active objects tend to have lower one-component retrieved $\log g$, suggesting that magnetic activity can bias simple spectral retrievals of late-M dwarf photospheres.
    }
    \label{fig:flare_and_X-ray-vs-retrieved_logg}
\end{figure*}
\begin{figure*}
    \centering
    \includegraphics[width=\textwidth, height=0.4\textheight, keepaspectratio]{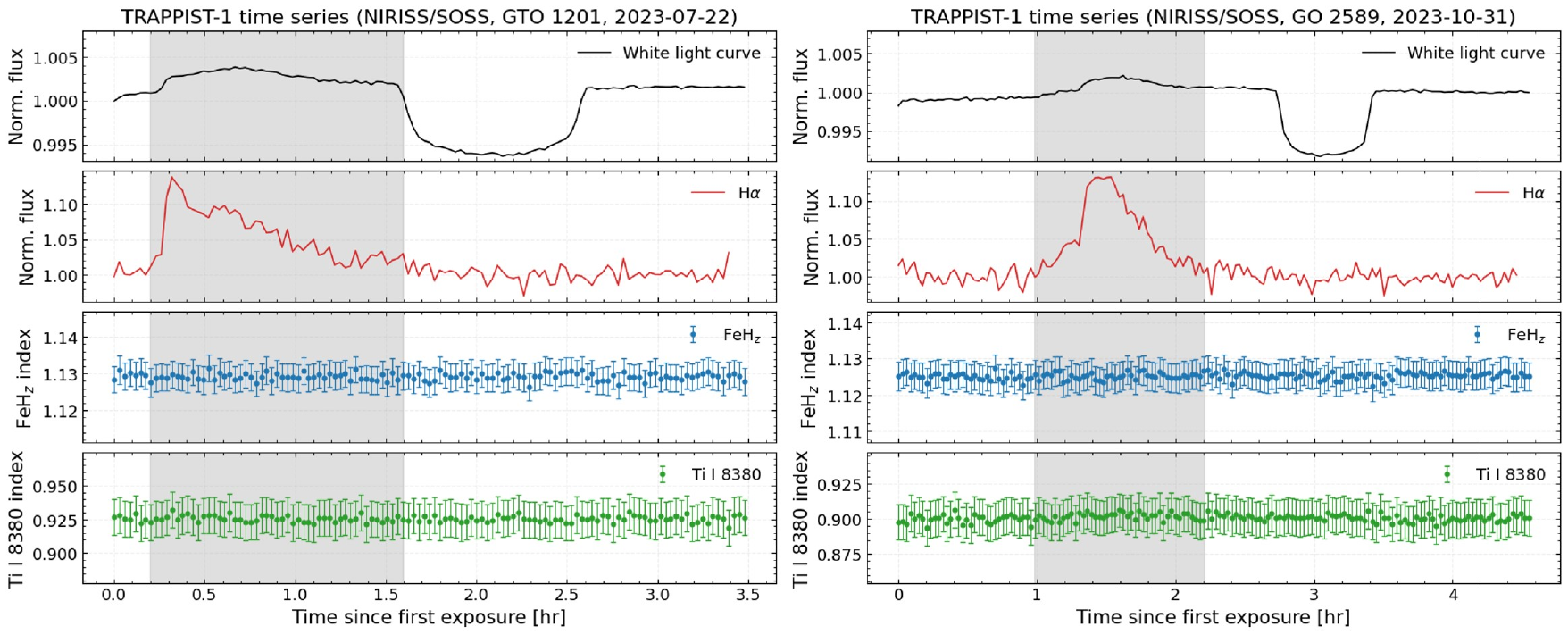}
    \caption{Time series of the white-light curve, H$\alpha$ light curve, FeH$_z$ index, and Ti~I~8380 index during the two most energetic flares observed in the TRAPPIST-1 NIRISS/SOSS visits.
    The shaded intervals mark the flaring phases determined from the H$\alpha$ light curves.
    The left panels show the flare observed during the GTO 1201 visit on July 22, 2023, and the right panels show the flare observed during the GO 2589 visit on October 31, 2023. The time axis is given in hours from the first exposure of each visit. Both flares show clear enhancements in the white-light and H$\alpha$ light curves, while no clear short-term variation is observed in the FeH$_z$ and Ti~\text{I}~8380 indices during the flare events.}
    \label{fig:trappist1_niriss_soss_flares_and_FeH}
\end{figure*}
\begin{figure}
    \centering
    \includegraphics[width=\textwidth, height=0.71\textheight, keepaspectratio]{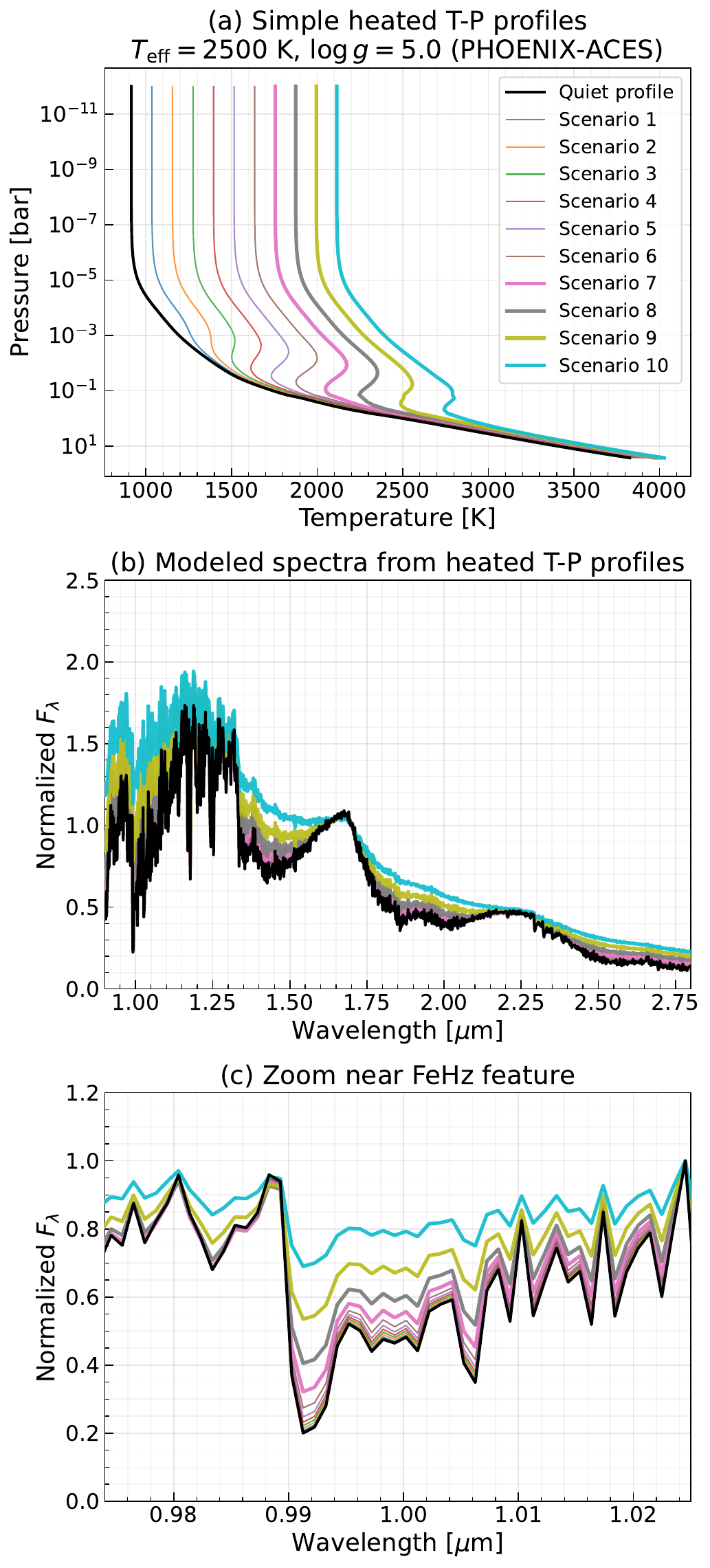}
    \caption{Simplified test of how imposed atmospheric heating changes the modeled near-infrared spectrum. Panel (a) shows the quiet PHOENIX-ACES T--P profile for $T_{\rm eff}=2500~{\rm K}$ and $\log g=5.0$, together with ten modified T--P profiles. 
    The heating scenarios are ordered from weak to strong, with Scenario~1 representing the weakest imposed heating and Scenario~10 representing the strongest imposed heating. Panel (b) shows the corresponding modeled spectra computed with the simplified \texttt{petitRADTRANS} setup and the fixed-VMR prescription described in Section~\ref{subsec: A toy moodel to test magnetic heating}. 
    The spectra are normalized for shape comparison. The broad H$_2$O absorption regions, especially around $1.3$--$1.5~\mu{\rm m}$, $1.75$--$2.05~\mu{\rm m}$, and $2.3$--$2.8~\mu{\rm m}$, become shallower for the stronger heating scenarios. 
    Panel (c) zooms in on the FeH$_z$ region near $0.998~\mu{\rm m}$, showing that this gravity-sensitive feature also responds noticeably to the imposed change in the thermal structure. 
    The thicker linewidths used for Scenarios 7--10 in Panel (a) highlight the cases that produce noticeable differences in the emergent spectra in Panel (b) and in the FeH$_z$ feature in Panel (c).
    The corresponding gravity-sensitive spectral index measurements (FeH$_z$, FeH$_J$, K~I$_J$, VO$_z$, and H-cont) for these spectra are shown in Figure~\ref{fig:gravity_sensitive_indices_of_different_heated_spectra}.}
    \label{fig:simple_modified_T-P_magnetic_heat_atmosphere_spectra}
\end{figure}
\begin{figure}
    \centering
    \includegraphics[width=\textwidth, height=0.5\textheight, keepaspectratio]{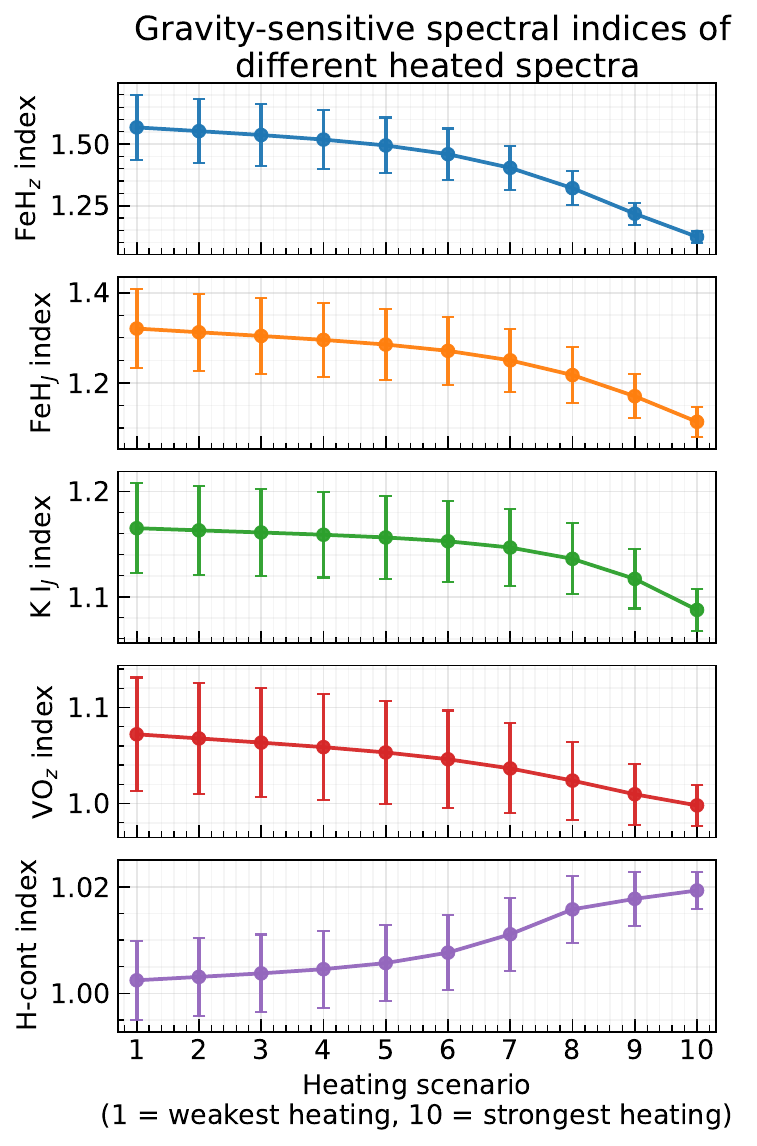}
    \caption{The five gravity-sensitive spectral indices measured from the modeled spectra of the ten heated-atmosphere scenarios. 
    The heating scenario number increases from the weakest imposed heating to the strongest imposed heating. 
    The FeH$_z$, FeH$_J$, K~I$_J$, and VO$_z$ indices decrease systematically as the imposed heating becomes stronger, while the H-cont index shows a slight increase. 
    Since the surface gravity is fixed in all scenarios, these variations are not caused by changes in $\log g$, but instead reflect the response of the gravity-sensitive spectral indices to the modified atmospheric thermal structure in this simple heating experiment.}
    \label{fig:gravity_sensitive_indices_of_different_heated_spectra}
\end{figure}

\begin{enumerate}
\item The free-$\log g$ scenario is favored for all late-M dwarfs in our sample (see Table~\ref{tab:retrieved_params_free_fixed_logg}, Figure~\ref{fig:1-component retrieval Trappist-1}, and Figure~\ref{fig:appendix_spex_retrievals}).
In this scenario, the retrieved surface gravities are all lower than the expected high-gravity values of late-M dwarfs, suggesting that current state-of-the-art stellar atmosphere models tend to lead to systematically underestimated surface gravities for late-M~dwarfs. 

\item All measured gravity-sensitive spectral indices show trends with the freely retrieved $\log g$ values, although the statistical significance varies among the indices (see Figure~\ref{fig:free-logg-vs-gravity_spectral_index}). 
FeH$_z$, FeH$_J$, and K\,\textsc{i}$_J$ show statistically significant positive correlations with the freely retrieved $\log g$ values, with Pearson coefficients of $r=0.576$, $0.576$, and $0.460$, respectively, and consistent positive Spearman coefficients. The corresponding p-values are all $\leq 0.03$.
H-cont shows a statistically significant anti-correlation with retrieved $\log g$, with $r=-0.474$, $\rho=-0.450$, and both corresponding $p\leq0.032$. 
The VO$_z$ index also shows a moderate positive trend with Pearson coefficient $r=0.455, p=0.0381$ and Spearman coefficient $\rho=0.391, p=0.0797$, but its weaker Spearman significance suggests that this relation should be interpreted more cautiously. 
Still, these trends indicate that the low-$\log g$ solutions are tracing real, systematic connections between the near-infrared spectral morphology and surface gravity, rather than being only numerical artifacts of the retrieval.

\item We find strong anti-correlations between freely retrieved $\log g$ values and magnetic activity indicators, including flare luminosity and X-ray luminosity.
These correlations indicate that more active stars tend to have lower retrieved $\log g$ values (see Figure~\ref{fig:flare_and_X-ray-vs-retrieved_logg}). 
These anti-correlations are statistically significant: both the $\log g$--$\log L_{\rm flare, TESS}$ and $\log g$--$\log L_X$ relations have Pearson and Spearman correlation coefficients below $-0.9$, with p-values of $p \leq 1.0\times10^{-4}$. We determine empirical correlations for both the $\log g$--spectral-index relations and the $\log g$--activity relations. Based on these connections, we consider magnetic heating a plausible mechanism for explaining the activity-dependent low-$\log g$ behavior observed in this work.

\item To further investigate this possible physical origin, we examined the NIRISS/SOSS spectra of TRAPPIST-1 taken during large flare events. 
We do not find clear short-timescale variations in the magnetically sensitive FeH$_z$ and Ti\,\textsc{i}~8380 indices during the detected flare events (see Figure~\ref{fig:trappist1_niriss_soss_flares_and_FeH}). 
This suggests that individual energetic flares may not be the dominant driver of the altered gravity-sensitive spectral morphology. Persistent or unresolved small-scale magnetic activity, such as microflare heating, may play a more important role in modifying the atmospheric thermal structure.

\item Finally, we performed a simplified radiative-transfer experiment to test how magnetic-heating-like changes in the T--P profile affect gravity-sensitive spectral diagnostics (see Section~\ref{subsec: A toy moodel to test magnetic heating}). 
Starting from a PHOENIX ACES atmosphere with $T_{\rm eff}=2500~{\rm K}$, $\log g=5.0$, and $[{\rm Fe/H}]=0$, we imposed pressure-dependent heating on the atmospheric thermal structure and computed spectra using a fixed-VMR opacity prescription (see Figure~\ref{fig:simple_modified_T-P_magnetic_heat_atmosphere_spectra}). 
All spectra were computed at the same input surface gravity of $\log g=5.0$.
Nevertheless, the imposed thermal modifications systematically change the gravity-sensitive indices: FeH$_z$, FeH$_J$, K\,\textsc{i}$_J$, and VO$_z$ decrease as the imposed heating becomes stronger, while H-cont increases (see Figure~\ref{fig:gravity_sensitive_indices_of_different_heated_spectra}).
This demonstrates that changes in the atmospheric thermal structure alone can modify spectral features that are commonly interpreted as gravity-sensitive diagnostics, even without changing the true surface gravity.
These behaviors are consistent with the observed activity--$\log g$ trends in this study, where more magnetically active stars tend to have lower retrieved $\log g$ values and correspondingly lower FeH$_z$, FeH$_J$, K\,\textsc{i}$_J$, and VO$_z$ indices, but higher H-cont values.
\end{enumerate}

Overall, our results suggest that the underestimated surface gravities retrieved from the observed NIR spectra of late-M dwarfs may not necessarily indicate genuinely low surface gravity. 
Instead, they may reflect activity-modified atmospheres whose emergent spectra mimic lower-gravity spectral morphology.
{
Our results are specific to the late-M spectral types studied here. 
Whether the same activity-related behavior extends to earlier M dwarfs remains unclear and will require a similar analysis across a broader range of spectral types (see Section~\ref{subsec:earlier_m_dwarfs}). 
}

These results may also have broader implications for studies of late-M dwarfs and their planetary systems. 
If persistent magnetic activity can modify the atmospheric T--P structure and produce spectra that resemble those of lower-gravity atmospheres, then the observationally quiescent spectra of active late-M dwarfs may still include the time-averaged effects of unresolved magnetic heating. 
In this picture, the photospheres of late-M dwarfs may be more dynamic than assumed in standard non-magnetic atmosphere models, which could complicate the interpretation of stellar spectra and exoplanet transmission spectra \citep[e.g.,][]{2018ApJ...853..122R, Rackham2019, Rackham2023, 2024AJ....168...82R}. 

Our results also suggest that gravity-sensitive spectral features may not be pure surface-gravity diagnostics for magnetically active very late-type dwarfs.
This has potential implications for age estimates of late-M dwarfs and early brown dwarfs before the L/T transition, because low-gravity spectral morphology is often used as an indicator of youth \citep[e.g.,][]{2004ApJ...600.1020M,2009AJ....137.3345C,2013ApJ...772...79A,2020MNRAS.491.5925M}. 
The age classifications of late-M dwarfs based solely on these gravity-sensitive features may be biased.
Flares and other magnetic-activity signatures have been detected in some early L-type brown dwarfs \citep[e.g.,][]{2013ApJ...779..172G,2016ApJ...828L..22S,2020MNRAS.494.5751P}, showing that early-L dwarfs can also be magnetically active.
Such objects may as well host smaller-scale magnetic activity that remains below current detection limits.
Therefore, if similar activity-associated atmospheric effects are present in early-L dwarfs, their age or youth classifications based only on gravity-sensitive spectral diagnostics may be biased too.
At the same time, these features may carry information about magnetic activity, providing an additional possible way to identify relatively quiet M dwarfs that are more favorable targets for precise exoplanet atmospheric characterization with current and future facilities, including JWST, Ariel \citep[e.g.,][]{Tinetti2018,Edwards2019}, and the proposed Nautilus Space Observatory \citep{2019AJ....158...83A}.
Future work should test this possibility with larger samples and simultaneous activity monitoring, especially with facilities designed to characterize the time-dependent activity of low-mass stars, such as SPARCS \citep[e.g.,][]{Ramiaramanantsoa2022, Shkolnik2025} and Pandora \citep[e.g.,][]{Quintana2021, Barclay2025, Rackham2026, Rotman2026}.

In addition, more physically self-consistent magnetic-heating atmosphere models will be needed to address the broader continuum discrepancy seen in the spectral modeling. 
Specifically, in this work, our simplified, toy experiment only modified the T--P profile and tested the response of local gravity-sensitive spectral features, without attempting to reproduce the full observed spectra or improve the continuum-level fit. 
A self-consistent treatment of the atmospheric structure, chemistry, opacities, and radiative transfer will be necessary to test whether magnetic heating can explain both the local spectral feature changes and the broader pseudo-continuum discrepancies in late-M dwarfs.

\begin{acknowledgments}
We thank the anonymous referee for their constructive comments, which helped
improve the clarity and impact of our study.
Based on observations with the NASA/ESA/CSA James Webb Space Telescope obtained from the Mikulski Archive for Space Telescopes (MAST) at the Space Telescope Science Institute, which is operated by the Association of Universities for Research in Astronomy, Incorporated, under NASA contract NAS5-03127. Support for program number JWST-AR-05370 was provided through a grant from the STScI under NASA contract NAS5-03127.
This material is based upon work supported by the National Aeronautics and Space Administration under Agreement No.\ 80NSSC21K0593 for the program ``Alien Earths.''
The results reported herein benefited from collaborations and/or information exchange within NASA's Nexus for Exoplanet System Science (NExSS) research coordination network sponsored by NASA's Science Mission Directorate.
This material is based upon work supported by the European Research Council (ERC) Synergy Grant under the European Union’s Horizon 2020 research and innovation program (grant No.\ 101118581---project REVEAL).
Visiting Astronomer at the Infrared Telescope Facility, which is operated by the University of Hawaii under contract 80HQTR24DA010 with the National Aeronautics and Space Administration.
\end{acknowledgments}

\clearpage

\begin{figure*}
    \centering
    \includegraphics[width=\textwidth, height=\textheight, keepaspectratio]{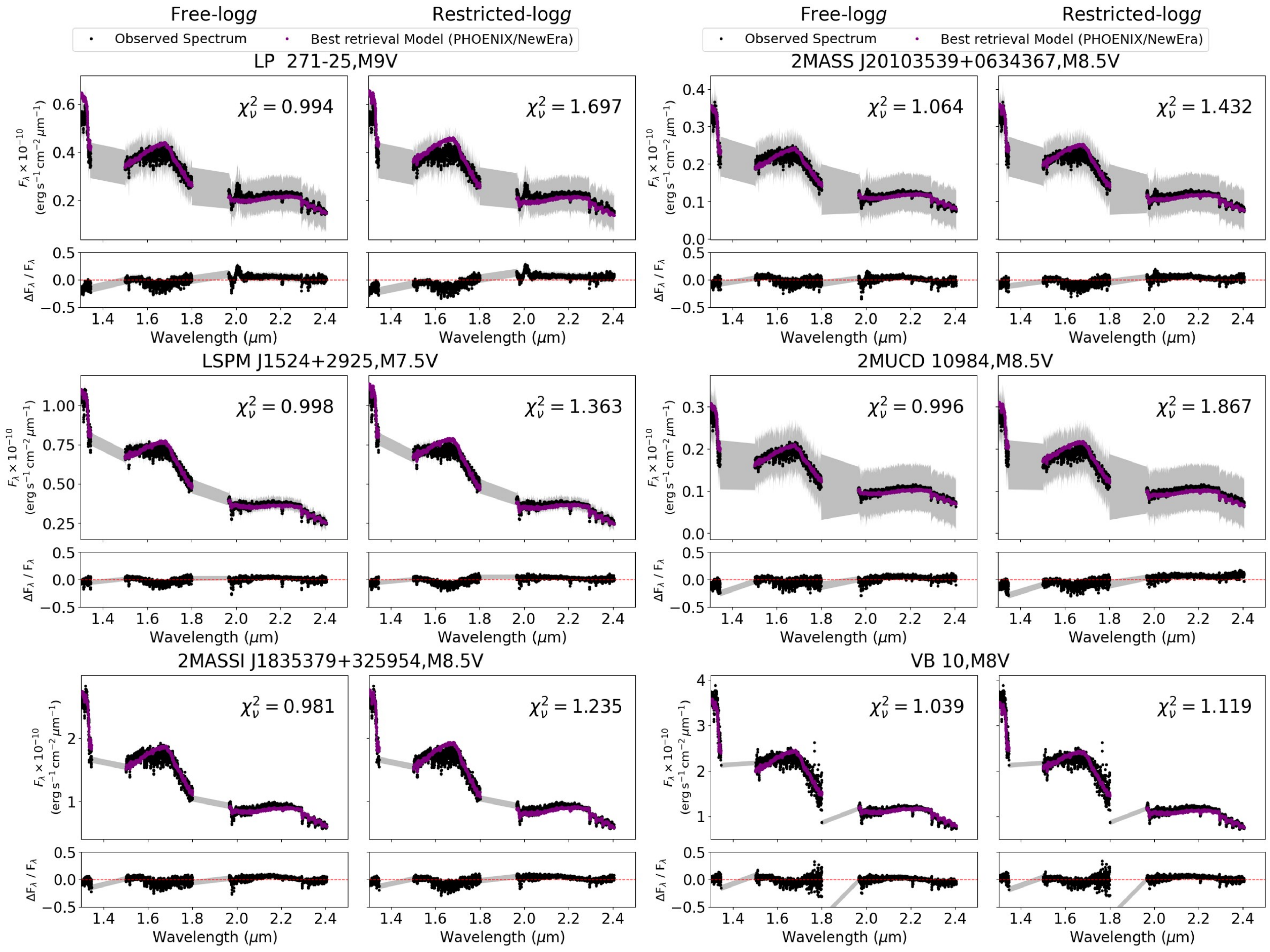}
    \caption{
    One-component spectral modeling of {six of the 22 M~dwarfs} with IRTF/SpeX spectra analyzed in this study, covering the wavelength range from 1.3~$\mu$m to 2.4~$\mu$m. 
    The figure follows the same format as Figure~\ref{fig:1-component retrieval Trappist-1}, comparing the best-retrieved models from the free-$\log g$ (panel on the left) and restricted-$\log g$ (panel on the right) scenarios for each star. 
    {The gray shaded regions represent the $1\sigma$ uncertainties, including the fitted error-inflation term in addition to the photon-noise uncertainty.}
    The best-retrieved parameters for both scenarios, the reduced chi-square values ($\chi^{2}_{\nu}$), and the model comparison metric $\Delta{\rm BIC}_{\rm free-R}$ of all these stars are listed in Table~\ref{tab:retrieved_params_free_fixed_logg}. 
    The remaining panels for the rest of {16 IRTF/SpeX M~dwarf stars} in our sample are shown in Figure~\ref{fig:appendix_spex_retrievals}.
}
    \label{fig:one_component_retrieval_all_M_dwarfs}
\end{figure*}





\begin{deluxetable*}{l l c l c l c}
\tabletypesize{\scriptsize}
\tablecaption{SpeX/IRTF observation summary of late-M dwarfs in this study \label{tab:sample_properties}}
\tablehead{
\colhead{Target} &
\colhead{Obs. Date} &
\colhead{Sp. Type} &
\colhead{Sp. Type Ref.} &
\colhead{$T_{\rm eff,lit}$ (K)} &
\colhead{$T_{\rm eff}$ Ref.} &
\colhead{$d_{\rm Gaia}$ (pc)$^{a}$}
}
\startdata
LP~399-274                & 2021 Aug 30 & M6 V   & \citet{2025AJ....170..213D}      & 2810$^{c}$             & \citet{2013ApJS..208....9P}           & 20.903 \\
NLTT~30915                & 2022 Feb 11 & M6.5 V & \citet{2025AJ....170..213D}      & 2745$^{c}$             & \citet{2013ApJS..208....9P}           & 20.814 \\
2MASSI~J1424187-351432    & 2025 Feb 06 & M6.5 V & \citet{2003AJ....126.2421C}      & 2745$^{c}$             & \citet{2013ApJS..208....9P}           & 19.328 \\
LSPM~J1327+0149           & 2022 Feb 11 & M7 V   & \citet{2025AJ....170..213D}      & 2738             & \citet{Wang2022}           & 23.851 \\
LP~335-12                 & 2025 Jun 19 & M7 V   & \citet{2015ApJS..219...33G}      & 2800             & \citet{2015ApJS..219...33G}           & 12.142 \\
UPM~J1425+2540            & 2022 Apr 19 & M7 V   & \citet{Wang2022}                 & 2867             & \citet{Wang2022}           & 16.096 \\
LSPM~J1524+2925           & 2025 Feb 06 & M7.5 V & \citet{2011ApJS..197...19K}      & 2625                    & \citet{2003ApJ...583..451M}      & 13.078 \\
VB~10                     & 2021 Oct 19 & M8 V   & \citet{1991ApJS...77..417K}      & $2508^{+60}_{-63}$      & \citet{Pineda2021}               & 5.919  \\
LP~315-53                 & 2025 Feb 20 & M8 V   & \citet{2011ApJS..197...19K}      & 2490             & \citet{2020ApJS..247...11R}           & 20.151 \\
SIPS~J1607-0442           & 2025 May 18 & M8 V   & \citet{2002ApJ...575..484G}      & $2466\pm30$             & \citet{2014AJ....147...94D}      & 15.270 \\
LP~326-21                 & 2024 May 10 & M8 V   & \citet{2000AJ....120.1085G}      & 2559                    & \citet{Gautier2007}              & 14.781 \\
DENIS~J104517.2-260724    & 2025 Feb 06 & M8 V   & \citet{2002ApJ...575..484G}      & 2570$^{c}$             & \citet{2013ApJS..208....9P}           & 23.678 \\
LP~412-31$^{b}$           & 2003 Sep 20 & M8 V   & \citet{2011ApJS..197...19K}      & 2550                    & \citet{2003ApJ...583..451M}      & 14.648 \\
2MASSI~J1835379+325954    & 2025 May 18 & M8.5 V & \citet{2012AJ....144...99D}      & 2440            & \citet{2020ApJS..247...11R}           & 5.689  \\
2MASS~J20103539+0634367   & 2025 Sep 05 & M8.5 V & \citet{2012ApJ...760..152L}      & 2475$^{c}$             & \citet{2013ApJS..208....9P}           & 16.168 \\
2MUCD~10984               & 2022 Feb 11 & M8.5 V & \citet{2003AJ....126.2421C}      & 2475$^{c}$             & \citet{2013ApJS..208....9P}           & 18.412 \\
TVLM~513-46               & 2025 Feb 20 & M8.5 V & \citet{2002AJ....123.2002H}      & 2475$^{c}$                    & \citet{Gautier2007}              & 10.734 \\
LP~271-25                 & 2023 Aug 24 & M9 V   & \citet{Bouy2022}                 & 2475$^{c}$             & \citet{2013ApJS..208....9P}           & 10.969 \\
DENIS-P~J1048-3956$^{b}$ & 2000 Dec 09 & M9 V   & \citet{Rajpurohit2013}           & $2450\pm50$             & \citet{Stelzer2012}              & 4.045  \\
LHS~2065$^{b}$            & 2000 Dec 10 & M9 V   & \citet{2005PASP..117..676R}      & 2420                    & \citet{Gautier2007}              & 8.659  \\
2MASSW~J0109216+294925    & 2025 Aug 04 & M9.5 V & \citet{2000AJ....120.1085G}      & 2350$^{c}$             & \citet{2013ApJS..208....9P}           & 15.871 \\
LP~944-20$^{b}$           & 2001 Jan 24 & M9.5 V & \citet{Martin2002}               & 2400             &     \citet{2020ApJS..247...11R}           & 6.427  \\
\enddata
\tablenotetext{a}{Distances are measured from parallax provided by the Gaia EDR3 catalog \citep{2020yCat.1350....0G}.}
\tablenotetext{b}{The spectra of these stars are from the IRTF Spectral Library of M~dwarfs \citep{2005ApJ...623.1115C, 2009ApJS..185..289R}.}
\tablenotetext{c}{ For targets without a literature $T_{\rm eff}$ measurement, the listed value is estimated by linearly interpolating the spectral-type--$T_{\rm eff}$ sequence from the stellar parameter table of E.~Mamajek (version 2022.04.16; \citealt{2013ApJS..208....9P}). }
\end{deluxetable*}


\begin{table*}
\centering
\caption{Observation summary of TRAPPIST-1 and excluded integration intervals for the nine TRAPPIST-1 NIRISS/SOSS visits. The masked fraction gives the percentage of each visit excluded due to planetary transits or stellar flares.}
\label{tab:excluded_integrations}
\begin{tabular}{ccccc}
\hline
Program & Target & Date & Excluded integrations & Masked fraction (\%) \\
\hline
GO~2589  & TRAPPIST-1~b & 2022 July 18    & 85--115 & 20.26 \\
GO~2589  & TRAPPIST-1~b & 2022 July 20    & 85--115 & 20.26 \\
GTO~1201 & TRAPPIST-1~f & 2022 October 28 & 20--105 & 71.07 \\
GO~2589  & TRAPPIST-1~c & 2022 October 28 & 85--120 & 22.78 \\
GTO~1201 & TRAPPIST-1~f & 2023 June 15    & 60--110 & 42.15 \\
GTO~1201 & TRAPPIST-1~f & 2023 June 24    & 45--85  & 33.88 \\
GTO~1201 & TRAPPIST-1~f & 2023 July 3     & 40--105 & 54.55 \\
GTO~1201 & TRAPPIST-1~f & 2023 July 22    & 7--90   & 69.42 \\
GO~2589  & TRAPPIST-1~c & 2023 October 31 & 20--75 \& 95--120 & 51.27 \\
\hline
\multicolumn{3}{c}{Total} & 14.66 hr of 35.49 hr excluded & 41.32 \\
\hline
\end{tabular}
\end{table*}

\begin{deluxetable*}{l c l}
\tabletypesize{\small}
\tablecaption{Adopted Stellar Properties of TRAPPIST-1\label{tab:trappist1_adopted_parameters}}
\tablehead{
\colhead{Parameters} &
\colhead{Value} &
\colhead{Reference}
}
\startdata
Spectral type & M8.0~V & \citet{2017Natur.542..456G} \\
$T_{\rm eff}$ & $2566 \pm 26~{\rm K}$ & \citet{2021PSJ.....2....1A} \\
$\mathrm{[Fe/H]}$ & $0.052 \pm 0.073$ & \citet{2024ApJ...970L...4D} \\
$M_\star$ & $0.0898 \pm 0.0023~M_\odot$ & \citet{2019ApJ...871...63M} \\
$R_\star$ & $0.1192 \pm 0.0013~R_\odot$ & \citet{2021PSJ.....2....1A} \\
$\log_{10} g_\star$ & $5.2396^{+0.0056}_{-0.0073}$ & \citet{2021PSJ.....2....1A} \\
\enddata
\tablecomments{
The surface gravity is given in cgs units of ${\rm cm~s^{-2}}$.
}
\end{deluxetable*}

\begin{table*}
\centering
\caption{Parameter settings for the one-component spectral modeling.}
\label{tab:one_component_priors}
\begin{tabular}{lcccc}
\hline
Scenario & $T_{\rm phot}$ & [Fe/H] & $\log g$ & Other parameters \\
\hline
Free-$\log g$ fit 
& Uniform [2300, 3500] K 
& Uniform [-1.0, 0.5] 
& Uniform [2.5, 5.5] 
& Truncated Gaussian priors for $R_\star$ and $d$ \\

Restricted-$\log g$ fit 
& Uniform [2300, 3500] K 
& Uniform [-1.0, 0.5] 
& Uniform [5.2, 5.3] 
& Truncated Gaussian priors for $R_\star$ and $d$ \\
\hline
\end{tabular}
\tablecomments{
{
For the stellar radius, we adopted truncated Gaussian priors centered on the literature value for TRAPPIST-1, with the standard deviation set to 40\% of that value and truncation bounds of $[0.05,\,0.2]\,R_{\odot}$.
} 
This interval was chosen to span the plausible radius range of the objects in our sample. 
For the distance, we adopted truncated Gaussian priors centered on the literature distance of each target listed in Table~\ref{tab:sample_properties}, with boundaries set to $\pm 10\%$ of the corresponding literature value.
}
\end{table*}

\begin{deluxetable}{lcccc}
\tablecaption{Adopted Band Definitions for the Gravity-sensitive Spectral Indices
\label{tab:gravity_index_bands}}
\tablewidth{0pt}
\tablehead{
\colhead{Index} &
\colhead{$\lambda_{\rm line}$} &
\colhead{$\lambda_{\rm cont1}$} &
\colhead{$\lambda_{\rm cont2}$} &
\colhead{Bandwidth} \\
\colhead{} &
\colhead{($\mu$m)} &
\colhead{($\mu$m)} &
\colhead{($\mu$m)} &
\colhead{($\mu$m)}
}
\startdata
FeH$_z$   & 0.998 & 0.980 & 1.022 & 0.0133 \\
VO$_z$    & 1.058 & 1.035 & 1.087 & 0.0141 \\
FeH$_J$   & 1.200 & 1.192 & 1.208 & 0.0024 \\
K~I$_J$   & 1.244 & 1.220 & 1.270 & 0.0166 \\
H-cont    & 1.560 & 1.270 & 1.670 & 0.0208
\enddata
\tablecomments{
The listed bandwidth is applied to the line band and both continuum bands for each index.
The band definitions for FeH$_z$, VO$_z$, FeH$_J$, and K~I$_J$ were adopted from Table~4 of \citet{2013ApJ...772...79A}.
For H-cont, we followed the modified definition proposed by \citet{2025AJ....170..213D}, adopting a blue-continuum band centered at 1.270~$\mu$m.
}
\end{deluxetable}

\begin{deluxetable*}{l c c c c c}
\tabletypesize{\scriptsize}
\tablecaption{TESS-band flare energies of TRAPPIST-1 from K2 and JWST NIRISS/SOSS observations
\label{tab:trappist1_flare_energies}}
\tablehead{
\colhead{Visit} &
\colhead{Flare No.} &
\colhead{Start} &
\colhead{Peak} &
\colhead{End} &
\colhead{$E_{\rm flare,TESS}$} \\
\colhead{} &
\colhead{} &
\colhead{(MJD)} &
\colhead{(MJD)} &
\colhead{(MJD)} &
\colhead{(erg)}
}
\startdata
K2 C12 & 1  & 57739.7765 & 57739.7799 & 57739.7813 & $3.298\times10^{30}$ \\
K2 C12 & 2  & 57744.6176 & 57744.6182 & 57744.6298 & $1.781\times10^{31}$ \\
K2 C12 & 3  & 57745.3987 & 57745.4008 & 57745.4049 & $2.807\times10^{30}$ \\
K2 C12 & 4  & 57752.3721 & 57752.3735 & 57752.3762 & $1.532\times10^{30}$ \\
K2 C12 & 5  & 57752.3817 & 57752.3851 & 57752.3933 & $8.318\times10^{30}$ \\
K2 C12 & 6  & 57752.3946 & 57752.3967 & 57752.3980 & $9.592\times10^{29}$ \\
K2 C12 & 7  & 57754.5733 & 57754.5747 & 57754.5767 & $1.576\times10^{30}$ \\
K2 C12 & 8  & 57755.4696 & 57755.4717 & 57755.4730 & $1.012\times10^{30}$ \\
K2 C12 & 9  & 57756.2569 & 57756.2576 & 57756.2617 & $1.790\times10^{30}$ \\
K2 C12 & 10 & 57757.4243 & 57757.4263 & 57757.4277 & $7.200\times10^{29}$ \\
K2 C12 & 11 & 57759.4293 & 57759.4620 & 57759.4851 & $4.696\times10^{31}$ \\
K2 C12 & 12 & 57761.8402 & 57761.8416 & 57761.8436 & $7.961\times10^{29}$ \\
K2 C12 & 13 & 57766.9951 & 57766.9958 & 57767.0053 & $1.167\times10^{31}$ \\
K2 C12 & 14 & 57767.0047 & 57767.0053 & 57767.0087 & $1.090\times10^{30}$ \\
K2 C12 & 15 & 57770.7872 & 57770.7886 & 57770.7947 & $8.886\times10^{30}$ \\
K2 C12 & 16 & 57771.6311 & 57771.6331 & 57771.6372 & $1.715\times10^{30}$ \\
K2 C12 & 17 & 57772.9407 & 57772.9421 & 57772.9666 & $2.413\times10^{31}$ \\
K2 C12 & 18 & 57773.8084 & 57773.8315 & 57773.8418 & $3.315\times10^{31}$ \\
K2 C12 & 19 & 57775.9911 & 57775.9918 & 57776.0014 & $1.449\times10^{31}$ \\
K2 C12 & 20 & 57783.4704 & 57783.4718 & 57783.4766 & $3.566\times10^{30}$ \\
K2 C12 & 21 & 57784.7120 & 57784.7283 & 57784.7379 & $2.406\times10^{31}$ \\
K2 C12 & 22 & 57784.8802 & 57784.8823 & 57784.8843 & $6.288\times10^{30}$ \\
K2 C12 & 23 & 57792.8553 & 57792.8560 & 57792.8587 & $1.209\times10^{30}$ \\
K2 C12 & 24 & 57793.5670 & 57793.5888 & 57793.5943 & $2.041\times10^{31}$ \\
K2 C12 & 25 & 57793.9484 & 57793.9518 & 57793.9518 & $1.087\times10^{30}$ \\
K2 C12 & 26 & 57799.2293 & 57799.2306 & 57799.2334 & $1.321\times10^{30}$ \\
K2 C12 & 27 & 57800.0036 & 57800.0043 & 57800.0070 & $1.572\times10^{30}$ \\
K2 C12 & 28 & 57800.1106 & 57800.1112 & 57800.1147 & $1.827\times10^{30}$ \\
K2 C12 & 29 & 57801.5837 & 57801.5844 & 57801.5871 & $1.528\times10^{30}$ \\
K2 C12 & 30 & 57802.1020 & 57802.1026 & 57802.1081 & $3.473\times10^{30}$ \\
K2 C12 & 31 & 57802.8729 & 57802.8736 & 57802.8899 & $1.729\times10^{31}$ \\
K2 C12 & 32 & 57802.8913 & 57802.8927 & 57802.8947 & $7.407\times10^{29}$ \\
K2 C12 & 33 & 57805.0332 & 57805.0346 & 57805.0366 & $5.790\times10^{30}$ \\
K2 C12 & 34 & 57805.8525 & 57805.8532 & 57805.8559 & $1.208\times10^{30}$ \\
K2 C12 & 35 & 57806.0085 & 57806.0112 & 57806.0303 & $4.071\times10^{31}$ \\
K2 C12 & 36 & 57812.1523 & 57812.1537 & 57812.1605 & $3.990\times10^{30}$ \\
K2 C12 & 37 & 57812.1598 & 57812.1612 & 57812.1639 & $1.097\times10^{30}$ \\
K2 C12 & 38 & 57812.3287 & 57812.3294 & 57812.3335 & $2.142\times10^{30}$ \\
K2 C12 & 39 & 57812.4411 & 57812.4445 & 57812.5058 & $4.666\times10^{32}$ \\
K2 C12 & 40 & 57812.5051 & 57812.5065 & 57812.5099 & $1.626\times10^{30}$ \\
K2 C12 & 41 & 57813.0847 & 57813.0860 & 57813.0881 & $1.902\times10^{31}$ \\
K2 C12 & 42 & 57814.3658 & 57814.3664 & 57814.3692 & $2.070\times10^{30}$ \\
K2 C12 & 43 & 57814.4645 & 57814.4659 & 57814.4686 & $8.762\times10^{30}$ \\
K2 C12 & 44 & 57816.1740 & 57816.1747 & 57816.1774 & $8.917\times10^{29}$ \\
\hline
GO 2589, 2022 Jul 20  & 1 & 59780.2339 & 59780.2375 & 59780.2520 & $4.032\times10^{29}$ \\
GTO 1201, 2022 Oct 28 & 2 & 59880.5036 & 59880.5072 & 59880.5133 & $6.907\times10^{29}$ \\
GTO 1201, 2023 Jun 15 & 3 & 60110.5890 & 60110.5939 & 60110.5975 & $2.430\times10^{29}$ \\
GTO 1201, 2023 Jul 03 & 4 & 60129.0542 & 60129.0591 & 60129.0736 & $1.990\times10^{30}$ \\
GTO 1201, 2023 Jul 22 & 5 & 60147.4266 & 60147.4314 & 60147.4688 & $2.809\times10^{30}$ \\
GO 2589, 2023 Oct 31  & 6 & 60248.9339 & 60248.9484 & 60249.0003 & $4.134\times10^{30}$ \\
\enddata
\tablecomments{
The first 44 flares were detected in the K2 Campaign 12 short-cadence observations, and the final six flares were detected in the JWST NIRISS/SOSS observations. 
The K2 times were converted from BKJD to MJD. All flare energies are expressed in the TESS bandpass.
}
\end{deluxetable*}

\begin{deluxetable*}{lcccc}
\tablecaption{Flare and X-ray Luminosities of the Late-M Dwarf Sample
\label{tab:flare_xray_luminosities}}
\tablehead{
\colhead{ID} &
\colhead{$L_{\rm flare, TESS}$} &
\colhead{$N_{\rm flare}$} &
\colhead{$L_X$} &
\colhead{$L_X$ Source} \\
\colhead{} &
\colhead{(${\rm erg~s^{-1}}$)} &
\colhead{} &
\colhead{(${\rm erg~s^{-1}}$)} &
\colhead{}
}
\startdata
LP~271-25 & \nodata & 0 & $<3.16\times10^{25}$ & \citet{1993ApJ...410..387F} \\
LSPM J1524+2925 & $(4.46\pm0.02)\times10^{24}$ & 3 & \nodata & \nodata \\
2MASSI J1835379+325954 & \nodata & 0 & $<3.30\times10^{24}$ & \citet{2023Natur.619..272K} \\
VB 10 & \nodata & \nodata & $(1.58\pm0.08)\times10^{25}$ & \citet{1999AJ....118.1369K} \\
TVLM 513-46 & \nodata & 0 & $(7.94\pm2.07)\times10^{24}$ & \citet{2008ApJ...673.1080B} \\
LP~315-53 & $(5.43\pm0.05)\times10^{26}$ & 6 & \nodata & \nodata \\
LP~326-21 & $(5.97\pm0.05)\times10^{25}$ & 2 & \nodata & \nodata \\
LP~412-31 & $(2.56\pm0.02)\times10^{26}$ & 11 & $(1.60\pm0.30)\times10^{27}$ & \citet{2006AandA...460L..35S} \\
DENIS-PJ1048-3956 & $(3.66\pm0.01)\times10^{23}$ & 4 & $(1.70\pm0.70)\times10^{25}$ & \citet{2012AandA...537A..94S} \\
LHS~2065 & $(3.26\pm0.01)\times10^{25}$ & 19 & $(2.20\pm0.70)\times10^{26}$ & \citet{2002AandA...382L...9S} \\
LP~944-20 & \nodata & \nodata & $<3.00\times10^{23}$ & \citet{2002NewA....7..595M} \\
TRAPPIST-1 & $(1.26\pm0.01)\times10^{26}$ & 50 & $(1.19\pm0.40)\times10^{27}$ & \citet{2017MNRAS.465L..74W} \\
LSPM J1327+0149 & $(2.23\pm0.01)\times10^{25}$ & 2 & \nodata & \nodata \\
LP~335-12 & $(6.70\pm0.03)\times10^{24}$ & 4 & \nodata & \nodata \\
NLTT 30915 & $(3.58\pm0.02)\times10^{25}$ & 4 & \nodata & \nodata \\
UPM J1425+2540 & $(3.78\pm0.02)\times10^{26}$ & 15 & \nodata & \nodata \\
LP~399-274 & $(7.59\pm0.04)\times10^{25}$ & 3 & \nodata & \nodata \\
2MASSI J1424187-351432 & $(1.03\pm0.01)\times10^{26}$ & 5 & \nodata & \nodata \\
\enddata
\tablecomments{
The flare luminosities and flare numbers listed here were derived from flares detected by \citet{2024AJ....168..234L} for the late-M dwarfs in our sample, except for TRAPPIST-1, for which the flare measurements are obtained from K2 short-cadence observations and {nine visits of NIRISS/SOSS observations} in this study.
}
\end{deluxetable*}

\begin{deluxetable*}{l l cc cc cc cc c cc c} 
\tabletypesize{\scriptsize} 
\rotate
\tablecaption{Retrieved stellar parameters from the free-$\log g$ and restricted (R)-$\log g$ fitting scenarios.\label{tab:retrieved_params_free_fixed_logg}} 
\tablehead{ \colhead{Simbad ID} & \colhead{SpT} & \multicolumn{2}{c}{$T_{\rm eff}$ (K)} & \multicolumn{2}{c}{$R_\star$ ($R_\odot$)} & \multicolumn{2}{c}{$\log g$} & \multicolumn{2}{c}{[Fe/H]} & \colhead{$\ln f_{var}$} & \multicolumn{2}{c}{$\chi^2_\nu$} & \colhead{$\Delta{\rm BIC}_{\rm free-R}$} \\ \colhead{} & \colhead{} & \colhead{Free-$\log g$} & \colhead{R-$\log g$} & \colhead{Free-$\log g$} & \colhead{R-$\log g$} & \colhead{Free-$\log g$} & \colhead{R-$\log g$} & \colhead{Free-$\log g$} & \colhead{R-$\log g$} & \colhead{} & \colhead{Free-$\log g$} & \colhead{R-$\log g$} & \colhead{} } 
\startdata 
LP~399-274 & M6V & $2762^{+29}_{-40}$ & $2895^{+30}_{-40}$ & $0.118^{+0.031}_{-0.014}$ & $0.125^{+0.015}_{-0.018}$ & $3.981^{+0.063}_{-0.048}$ & $5.226^{+0.016}_{-0.019}$ & $0.290^{+0.088}_{-0.081}$ & $0.490^{+0.084}_{-0.069}$ & $-3.220^{+0.014}_{-0.017}$ & $0.979$ & $1.108$ & $-325.4$ \\ NLTT~30915 & M6.5V & $2805^{+30}_{-39}$ & $2870^{+35}_{-30}$ & $0.145^{+0.014}_{-0.009}$ & $0.116^{+0.018}_{-0.012}$ & $4.092^{+0.066}_{-0.075}$ & $5.202^{+0.021}_{-0.015}$ & $0.013^{+0.138}_{-0.187}$ & $0.302^{+0.088}_{-0.079}$ & $-3.382^{+0.012}_{-0.016}$ & $1.022$ & $2.248$ & $-3168.1$ \\ 2MASSI J1424187-351432 & M6.5V & $2705^{+33}_{-37}$ & $2743^{+34}_{-33}$ & $0.126^{+0.014}_{-0.013}$ & $0.119^{+0.009}_{-0.010}$ & $3.954^{+0.066}_{-0.066}$ & $5.220^{+0.016}_{-0.018}$ & $-0.176^{+0.191}_{-0.134}$ & $0.427^{+0.082}_{-0.090}$ & $-3.205^{+0.014}_{-0.015}$ & $1.029$ & $1.188$ & $-403.1$ \\ LSPM J1327+0149 & M7V & $2682^{+41}_{-29}$ & $2754^{+37}_{-32}$ & $0.116^{+0.015}_{-0.013}$ & $0.110^{+0.016}_{-0.015}$ & $4.297^{+0.015}_{-0.013}$ & $5.226^{+0.011}_{-0.016}$ & $0.073^{+0.188}_{-0.120}$ & $0.085^{+0.101}_{-0.086}$ & $-3.380^{+0.015}_{-0.014}$ & $0.966$ & $1.221$ & $-653.0$ \\ LP~335-12 & M7V & $2741^{+29}_{-34}$ & $2786^{+26}_{-43}$ & $0.122^{+0.011}_{-0.014}$ & $0.117^{+0.014}_{-0.014}$ & $4.304^{+0.167}_{-0.176}$ & $5.245^{+0.014}_{-0.016}$ & $0.333^{+0.091}_{-0.083}$ & $0.359^{+0.084}_{-0.077}$ & $-3.327^{+0.015}_{-0.014}$ & $0.983$ & $1.363$ & $-976.0$ \\ UPM J1425+2540 & M7V & $2809^{+38}_{-35}$ & $2872^{+33}_{-35}$ & $0.150^{+0.009}_{-0.013}$ & $0.127^{+0.012}_{-0.012}$ & $3.925^{+0.065}_{-0.072}$ & $5.217^{+0.018}_{-0.018}$ & $0.169^{+0.171}_{-0.179}$ & $0.296^{+0.090}_{-0.073}$ & $-3.369^{+0.014}_{-0.015}$ & $0.939$ & $1.050$ & $-280.6$ \\ LSPM J1524+2925 & M7.5V & $2666^{+16}_{-17}$ & $2696^{+32}_{-33}$ & $0.101^{+0.023}_{-0.014}$ & $0.118^{+0.014}_{-0.018}$ & $4.244^{+0.015}_{-0.015}$ & $5.244^{+0.017}_{-0.016}$ & $0.063^{+0.111}_{-0.164}$ & $0.339^{+0.085}_{-0.073}$ & $-3.244^{+0.013}_{-0.014}$ & $0.998$ & $1.363$ & $-939.4$ \\ VB~10 & M8V & $2566^{+15}_{-18}$ & $2711^{+22}_{-39}$ & $0.122^{+0.010}_{-0.012}$ & $0.122^{+0.015}_{-0.009}$ & $4.595^{+0.078}_{-0.054}$ & $5.231^{+0.016}_{-0.014}$ & $0.251^{+0.087}_{-0.080}$ & $0.256^{+0.091}_{-0.076}$ & $-2.963^{+0.015}_{-0.014}$ & $1.039$ & $1.119$ & $-198.0$ \\ LP~315-53 & M8V & $2683^{+16}_{-17}$ & $2785^{+31}_{-37}$ & $0.136^{+0.015}_{-0.012}$ & $0.124^{+0.013}_{-0.015}$ & $3.890^{+0.053}_{-0.045}$ & $5.234^{+0.015}_{-0.018}$ & $0.452^{+0.079}_{-0.077}$ & $0.323^{+0.076}_{-0.081}$ & $-3.148^{+0.013}_{-0.014}$ & $0.997$ & $1.387$ & $-1002.9$ \\ SIPS J1607-0442 & M8V & $2620^{+31}_{-31}$ & $2654^{+34}_{-34}$ & $0.116^{+0.012}_{-0.010}$ & $0.082^{+0.015}_{-0.014}$ & $4.030^{+0.087}_{-0.075}$ & $5.205^{+0.017}_{-0.018}$ & $0.404^{+0.086}_{-0.073}$ & $0.447^{+0.086}_{-0.089}$ & $-2.945^{+0.015}_{-0.015}$ & $0.977$ & $1.287$ & $-795.5$ \\ LP~326-21 & M8V & $2619^{+37}_{-32}$ & $2707^{+39}_{-33}$ & $0.123^{+0.011}_{-0.010}$ & $0.106^{+0.013}_{-0.013}$ & $4.131^{+0.091}_{-0.082}$ & $5.232^{+0.019}_{-0.017}$ & $0.391^{+0.085}_{-0.078}$ & $0.275^{+0.092}_{-0.077}$ & $-3.155^{+0.015}_{-0.013}$ & $1.043$ & $1.736$ & $-1788.1$ \\ DENIS J104517.2-260724 & M8V & $2633^{+32}_{-34}$ & $2705^{+33}_{-30}$ & $0.133^{+0.016}_{-0.012}$ & $0.110^{+0.011}_{-0.009}$ & $4.057^{+0.080}_{-0.077}$ & $5.223^{+0.014}_{-0.018}$ & $0.493^{+0.091}_{-0.091}$ & $0.273^{+0.081}_{-0.088}$ & $-2.831^{+0.014}_{-0.015}$ & $1.007$ & $1.386$ & $-973.5$ \\ LP412-31 & M8V & $2564^{+33}_{-34}$ & $2720^{+33}_{-35}$ & $0.119^{+0.001}_{-0.001}$ & $0.120^{+0.001}_{-0.001}$ & $3.885^{+0.174}_{-0.163}$ & $5.247^{+0.017}_{-0.017}$ & $0.332^{+0.065}_{-0.070}$ & $0.080^{+0.014}_{-0.014}$ & $-2.933^{+0.015}_{-0.014}$ & $0.993$ & $1.946$ & $-2398.5$ \\ TRAPPIST-1 & M8V & $2483^{+32}_{-34}$ & $2657^{+17}_{-17}$ & $0.120^{+0.001}_{-0.001}$ & $0.119^{+0.001}_{-0.001}$ & $3.995^{+0.222}_{-0.217}$ & $5.238^{+0.002}_{-0.002}$ & $0.351^{+0.074}_{-0.089}$ & $0.267^{+0.089}_{-0.083}$ & $-3.456^{+0.017}_{-0.018}$ & $1.019$ & $3.348$ & $-3576.9$ \\ 2MASSI J1835379+325954 & M8.5V & $2556^{+35}_{-27}$ & $2650^{+31}_{-33}$ & $0.102^{+0.009}_{-0.012}$ & $0.122^{+0.011}_{-0.013}$ & $4.478^{+0.070}_{-0.057}$ & $5.245^{+0.017}_{-0.016}$ & $0.394^{+0.063}_{-0.098}$ & $0.402^{+0.084}_{-0.085}$ & $-3.018^{+0.013}_{-0.013}$ & $0.981$ & $1.235$ & $-651.4$ \\ 2MASS J20103539+0634367 & M8.5V & $2540^{+31}_{-36}$ & $2594^{+35}_{-34}$ & $0.114^{+0.011}_{-0.011}$ & $0.107^{+0.009}_{-0.014}$ & $4.076^{+0.089}_{-0.077}$ & $5.217^{+0.018}_{-0.015}$ & $0.475^{+0.093}_{-0.077}$ & $0.449^{+0.088}_{-0.078}$ & $-3.008^{+0.015}_{-0.016}$ & $1.064$ & $1.432$ & $-944.9$ \\ 2MUCD 10984 & M8.5V & $2462^{+33}_{-35}$ & $2603^{+28}_{-41}$ & $0.107^{+0.011}_{-0.009}$ & $0.123^{+0.014}_{-0.016}$ & $4.102^{+0.090}_{-0.073}$ & $5.209^{+0.017}_{-0.018}$ & $0.301^{+0.083}_{-0.088}$ & $0.423^{+0.084}_{-0.098}$ & $-2.933^{+0.014}_{-0.015}$ & $0.996$ & $1.867$ & $-2248.9$ \\ TVLM~513-46 & M8.5V & $2453^{+30}_{-37}$ & $2563^{+25}_{-35}$ & $0.089^{+0.016}_{-0.012}$ & $0.092^{+0.011}_{-0.018}$ & $4.436^{+0.033}_{-0.032}$ & $5.205^{+0.011}_{-0.021}$ & $0.303^{+0.060}_{-0.081}$ & $0.276^{+0.099}_{-0.048}$ & $-2.795^{+0.012}_{-0.014}$ & $0.994$ & $1.353$ & $-921.5$ \\ LP~271-25 & M9V & $2491^{+46}_{-31}$ & $2603^{+34}_{-28}$ & $0.072^{+0.010}_{-0.018}$ & $0.093^{+0.012}_{-0.018}$ & $4.248^{+0.013}_{-0.021}$ & $5.230^{+0.017}_{-0.015}$ & $0.291^{+0.095}_{-0.061}$ & $0.268^{+0.061}_{-0.086}$ & $-2.625^{+0.013}_{-0.012}$ & $0.994$ & $1.697$ & $-1813.8$ \\ DENIS-PJ1048-3956 & M9V & $2452^{+34}_{-33}$ & $2614^{+33}_{-36}$ & $0.120^{+0.001}_{-0.001}$ & $0.120^{+0.001}_{-0.001}$ & $4.436^{+0.167}_{-0.169}$ & $5.227^{+0.017}_{-0.018}$ & $0.346^{+0.085}_{-0.087}$ & $0.391^{+0.086}_{-0.085}$ & $-3.014^{+0.014}_{-0.014}$ & $0.994$ & $1.718$ & $-1893.7$ \\ LHS2065 & M9V & $2474^{+34}_{-34}$ & $2619^{+33}_{-34}$ & $0.119^{+0.001}_{-0.001}$ & $0.121^{+0.001}_{-0.001}$ & $4.175^{+0.101}_{-0.103}$ & $5.228^{+0.017}_{-0.017}$ & $0.258^{+0.085}_{-0.083}$ & $0.296^{+0.083}_{-0.087}$ & $-2.453^{+0.014}_{-0.014}$ & $0.977$ & $1.483$ & $-1169.7$ \\ 2MASSW J0109216+294925 & M9.5V & $2514^{+17}_{-23}$ & $2637^{+32}_{-36}$ & $0.068^{+0.023}_{-0.001}$ & $0.099^{+0.008}_{-0.018}$ & $4.211^{+0.016}_{-0.086}$ & $5.248^{+0.016}_{-0.016}$ & $0.423^{+0.031}_{-0.043}$ & $0.405^{+0.079}_{-0.090}$ & $-2.822^{+0.008}_{-0.013}$ & $0.978$ & $1.655$ & $-1746.1$ \\ LP944-20 & M9.5V & $2483^{+33}_{-35}$ & $2585^{+33}_{-32}$ & $0.113^{+0.003}_{-0.003}$ & $0.100^{+0.003}_{-0.003}$ & $4.720^{+0.083}_{-0.086}$ & $5.247^{+0.017}_{-0.016}$ & $0.364^{+0.085}_{-0.086}$ & $0.457^{+0.085}_{-0.082}$ & $-2.765^{+0.014}_{-0.013}$ & $1.021$ & $1.909$ & $-2269.1$ \\ 
\enddata 
\tablecomments{Values are reported as best-fit parameters with asymmetric 16th--84th percentile posterior uncertainties. $\Delta{\rm BIC}_{\rm free-R}={\rm BIC}_{\rm free}-{\rm BIC}_{\rm R}$; negative values indicate that the free-$\log g$ scenario is preferred by BIC.} 
\end{deluxetable*}

\begin{deluxetable*}{lccccc}
\tablecaption{Gravity-sensitive spectral indices measured for the late-M dwarf sample\label{tab:gravity_sensitive_indices}}
\tablehead{
\colhead{Target} &
\colhead{FeH$_z$} &
\colhead{FeH$_J$} &
\colhead{K I$_J$} &
\colhead{VO$_z$} &
\colhead{H-cont}
}
\startdata
LP 399-274                 & $1.146 \pm 0.009$ & $1.067 \pm 0.010$ & $1.065 \pm 0.004$ & $0.997 \pm 0.006$ & $0.996 \pm 0.004$ \\
NLTT 30915                 & $1.154 \pm 0.009$ & $1.088 \pm 0.008$ & $1.058 \pm 0.004$ & $0.998 \pm 0.006$ & $0.987 \pm 0.004$ \\
2MASSI J1424187-351432     & $1.143 \pm 0.010$ & $1.122 \pm 0.012$ & $1.079 \pm 0.004$ & $1.008 \pm 0.007$ & $1.005 \pm 0.006$ \\
LSPM J1327+0149            & $1.176 \pm 0.010$ & $1.107 \pm 0.010$ & $1.071 \pm 0.005$ & $1.012 \pm 0.007$ & $0.990 \pm 0.004$ \\
LP 335-12                  & $1.208 \pm 0.011$ & $1.122 \pm 0.010$ & $1.078 \pm 0.005$ & $1.020 \pm 0.007$ & $0.988 \pm 0.004$ \\
UPM J1425+2540             & $1.122 \pm 0.007$ & $1.082 \pm 0.009$ & $1.055 \pm 0.004$ & $1.013 \pm 0.005$ & $0.992 \pm 0.003$ \\
LSPM J1524+2925            & $1.150 \pm 0.009$ & $1.117 \pm 0.012$ & $1.071 \pm 0.004$ & $1.032 \pm 0.007$ & $0.989 \pm 0.004$ \\
VB 10                      & $1.191 \pm 0.011$ & $1.150 \pm 0.010$ & $1.090 \pm 0.006$ & $1.072 \pm 0.008$ & $0.987 \pm 0.004$ \\
LP 315-53                  & $1.127 \pm 0.009$ & $1.105 \pm 0.011$ & $1.068 \pm 0.005$ & $1.042 \pm 0.007$ & $0.996 \pm 0.004$ \\
SIPS J1607-0442            & $1.174 \pm 0.010$ & $1.142 \pm 0.012$ & $1.087 \pm 0.006$ & $1.074 \pm 0.008$ & $0.985 \pm 0.004$ \\
LP 326-21                  & $1.191 \pm 0.011$ & $1.138 \pm 0.013$ & $1.095 \pm 0.005$ & $1.041 \pm 0.008$ & $0.982 \pm 0.004$ \\
DENIS J104517.2-260724     & $1.267 \pm 0.014$ & $1.170 \pm 0.013$ & $1.102 \pm 0.006$ & $1.067 \pm 0.009$ & $0.979 \pm 0.005$ \\
LP412-31                   & $1.205 \pm 0.017$ & $1.108 \pm 0.016$ & $1.101 \pm 0.006$ & $1.054 \pm 0.010$ & $0.980 \pm 0.005$ \\
TRAPPIST-1                 & $1.131 \pm 0.007$ & $1.078 \pm 0.006$ & $1.093 \pm 0.005$ & $1.075 \pm 0.006$ & $0.987 \pm 0.003$ \\
2MASSI J1835379+325954     & $1.338 \pm 0.018$ & $1.202 \pm 0.021$ & $1.129 \pm 0.007$ & $1.064 \pm 0.011$ & $0.972 \pm 0.005$ \\
2MASS J20103539+0634367    & $1.272 \pm 0.017$ & $1.206 \pm 0.021$ & $1.124 \pm 0.007$ & $1.061 \pm 0.011$ & $0.970 \pm 0.006$ \\
2MUCD 10984                & $1.270 \pm 0.016$ & $1.187 \pm 0.018$ & $1.120 \pm 0.007$ & $1.067 \pm 0.010$ & $0.973 \pm 0.006$ \\
TVLM 513-46                & $1.329 \pm 0.018$ & $1.226 \pm 0.022$ & $1.136 \pm 0.008$ & $1.093 \pm 0.012$ & $0.963 \pm 0.006$ \\
LP 271-25                  & $1.235 \pm 0.014$ & $1.192 \pm 0.022$ & $1.129 \pm 0.007$ & $1.143 \pm 0.011$ & $0.974 \pm 0.006$ \\
DENIS-P J1048-3956         & $1.300 \pm 0.019$ & $1.190 \pm 0.025$ & $1.124 \pm 0.008$ & $1.059 \pm 0.012$ & $0.977 \pm 0.007$ \\
LHS 2065                   & $1.242 \pm 0.014$ & $1.192 \pm 0.019$ & $1.111 \pm 0.008$ & $1.126 \pm 0.011$ & $0.980 \pm 0.006$ \\
2MASSW J0109216+294925     & $1.249 \pm 0.018$ & $1.188 \pm 0.020$ & $1.120 \pm 0.009$ & $1.154 \pm 0.013$ & $0.973 \pm 0.006$ \\
LP 944-20                  & $1.263 \pm 0.019$ & $1.197 \pm 0.025$ & $1.117 \pm 0.009$ & $1.121 \pm 0.013$ & $0.973 \pm 0.007$ \\
\enddata
\end{deluxetable*}

\clearpage
\appendix

\section{Remaining SpeX/IRTF Spectral Modeling Results}

\label{subsec:additional_spex_retrievals}

Figure~\ref{fig:appendix_spex_retrievals} shows the remaining SpeX/IRTF retrievals, following the format of Figure~\ref{fig:one_component_retrieval_all_M_dwarfs}.

\begin{figure*}
    \centering
    \figurenum{A.1}
    \includegraphics[width=\textwidth, height=0.6\textheight, keepaspectratio]{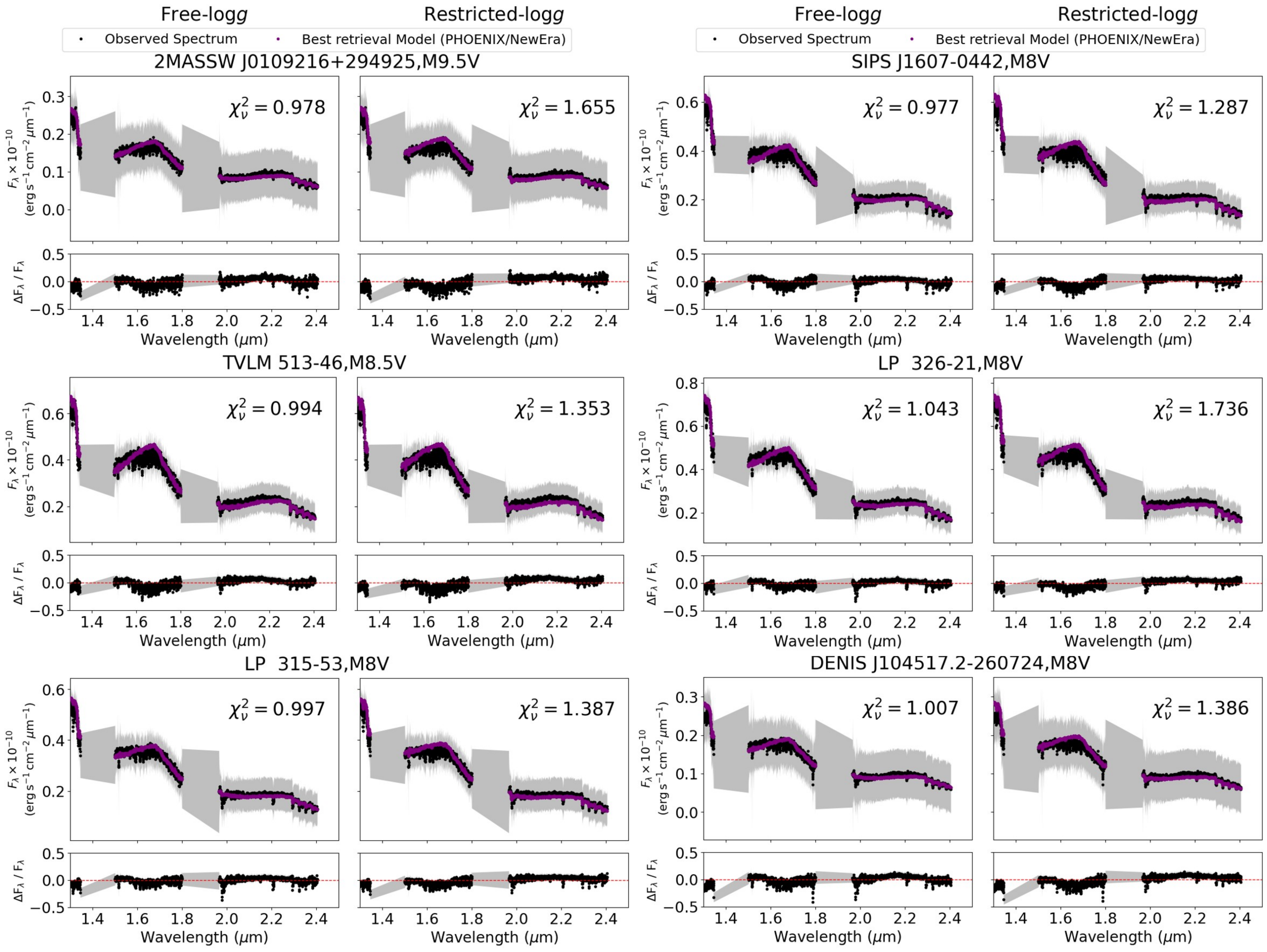}
    \caption{ 
    These panels show the remaining retrieval results for the {16 IRTF/SpeX} late-M~dwarfs not displayed in the main-text figure, following the same format as Figure~\ref{fig:one_component_retrieval_all_M_dwarfs}. 
    For each star, the left and right panels compare the best-fit retrieved models from the free-$\log g$ and restricted-$\log g$ scenarios, respectively. 
    {The gray shaded regions represent the $1\sigma$ uncertainties, including the fitted error-inflation term in addition to the photon-noise uncertainty.}
    The best-fit retrieved parameters, reduced chi-square values ($\chi^{2}_{\nu}$), and $\Delta{\rm BIC}_{\rm free-R}$ values are listed in Table~\ref{tab:retrieved_params_free_fixed_logg}.}
    \label{fig:appendix_spex_retrievals}
\end{figure*}

\begin{figure*}
    \centering
    \figurenum{A.1}
    \includegraphics[width=\textwidth, height=0.6\textheight, keepaspectratio]{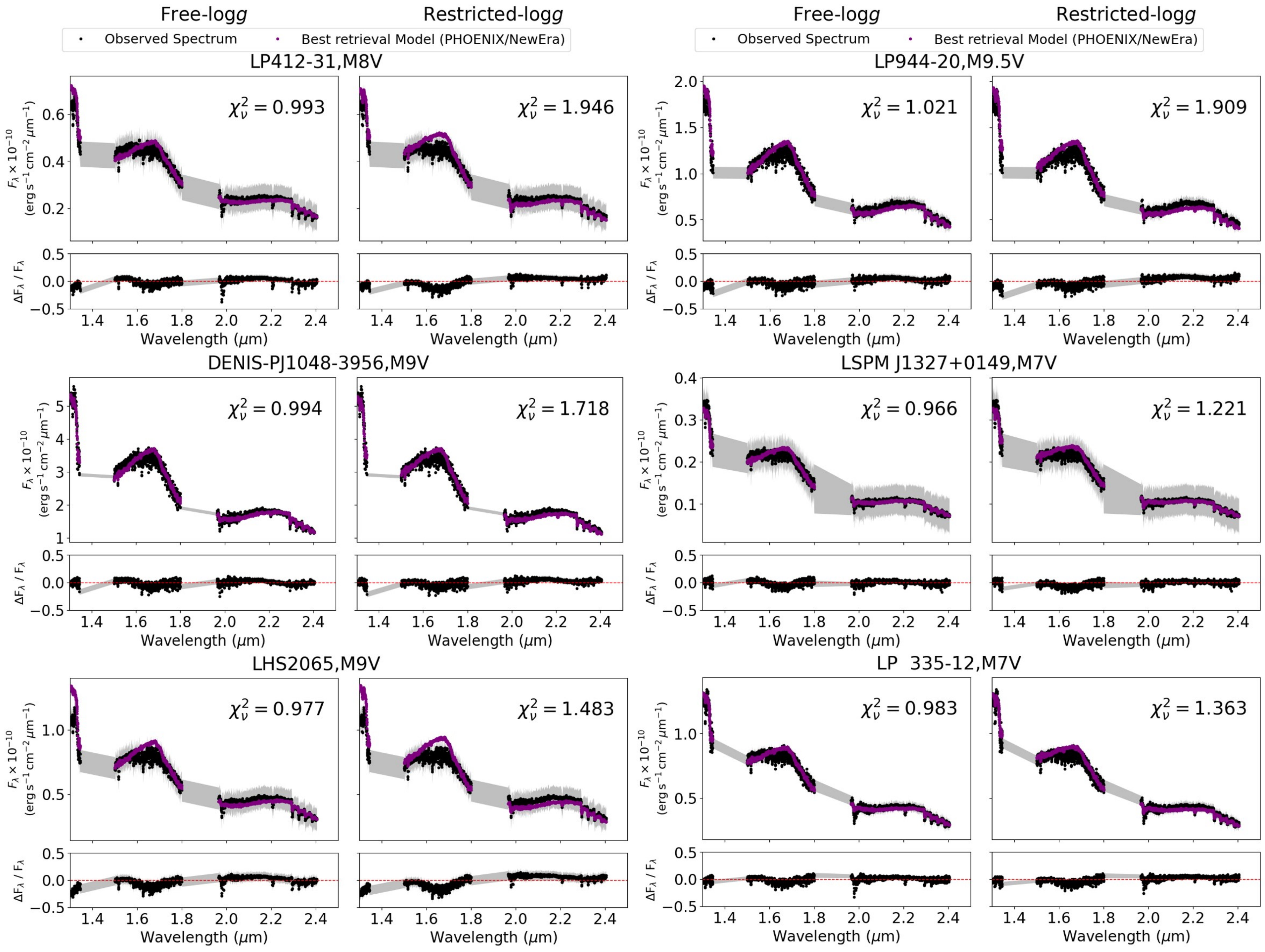}
    \caption{Continued.}
\end{figure*}

\begin{figure*}
    \centering
    \figurenum{A.1}
    \includegraphics[width=\textwidth, height=0.6\textheight, keepaspectratio]{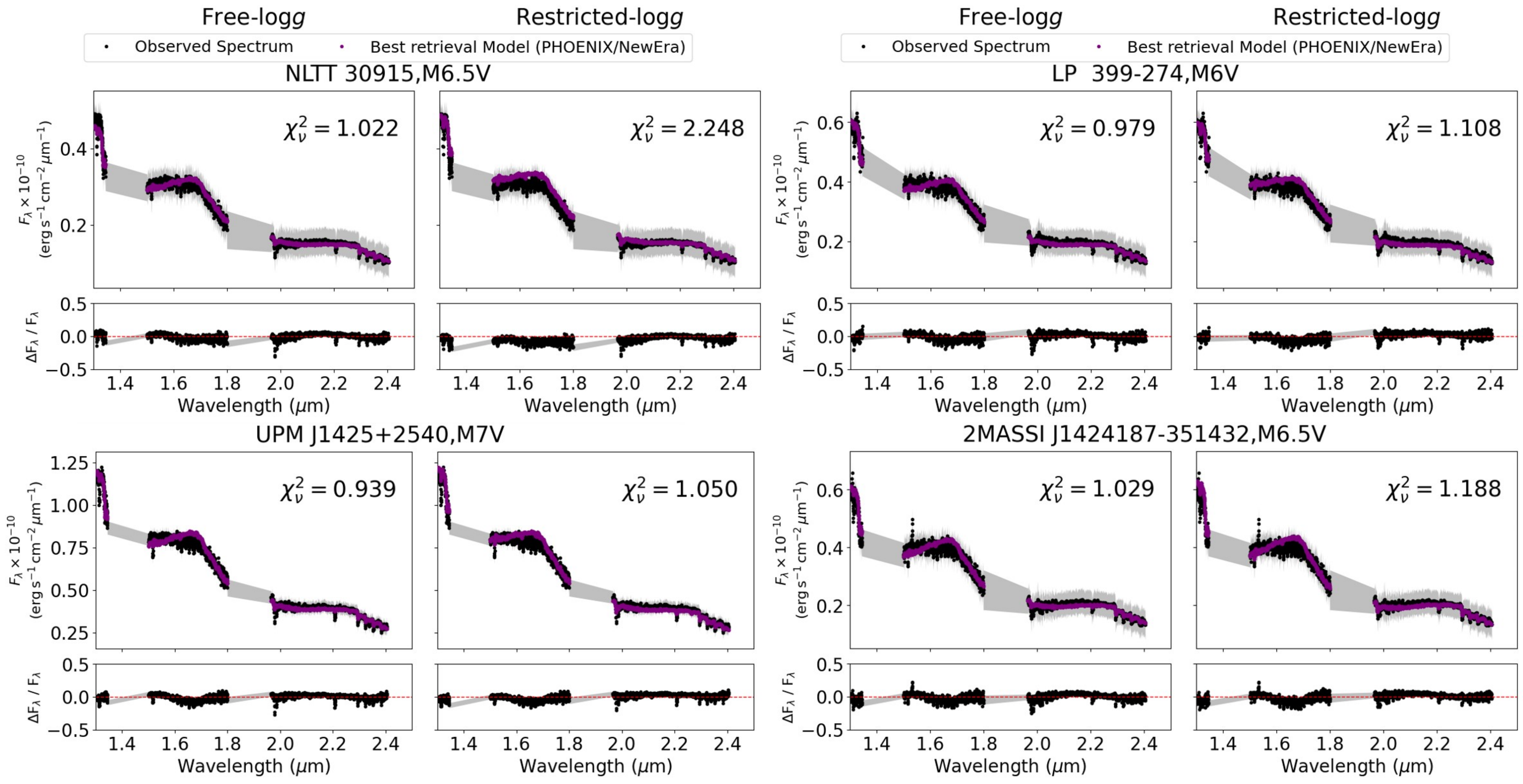}
    \caption{Continued.}
\end{figure*}

\clearpage

\section{{Testing Whether the Different Wavelength Coverage Affects the Comparability of the NIRISS/SOSS and IRTF/SpeX Retrievals}}
\label{subsec:wcoverage_test_for_retrievals}

\begin{figure}
    \centering
    \figurenum{A.2}
    \includegraphics[width=\textwidth, height=0.35\textheight, keepaspectratio]
    {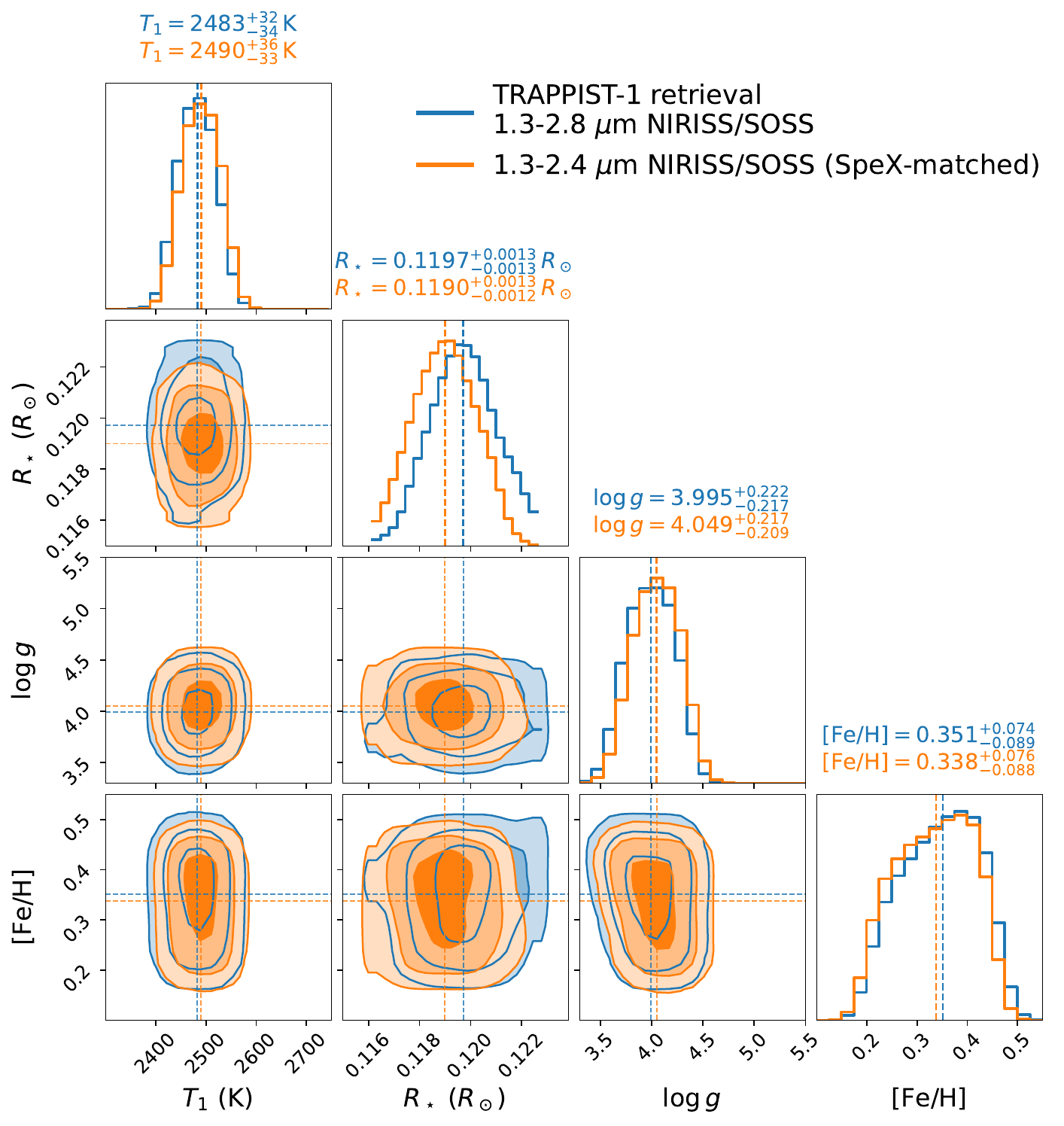}
    \caption{{ Comparison of the posterior distributions from the free-$\log g$ retrieval of TRAPPIST-1's NIRISS/SOSS spectrum using the wavelength range of 1.3--2.8~$\mu$m (blue) and the same spectrum restricted to the wavelength coverage used for the IRTF/SpeX sample (orange). 
    For the latter, the NIRISS/SOSS spectrum is trimmed to 1.3--2.4~$\mu$m, with the same regions affected by strong telluric absorption in the SpeX spectra excluded. 
    The dashed lines indicate the best-fit parameter values, and the values above the diagonal panels show the corresponding best-fit values with their 16th--84th percentile posterior uncertainties. The retrieved $T_1$ (or $T_{eff}$ ), $R_\star$, $\log g$, and [Fe/H] remain consistent within their uncertainties between the two retrievals. }}
    \label{fig:1_corner_plots_wcoverage_test}
\end{figure}

{
The NIRISS/SOSS spectrum of TRAPPIST-1 used in our spectral modeling covers 1.3--2.8~$\mu$m, while the IRTF/SpeX spectra are modeled over an overall wavelength range of 1.3--2.4~$\mu$m, with regions strongly affected by telluric water-vapor absorption excluded. 
To test whether this difference in wavelength coverage affects the retrieved stellar parameters, we repeated the free-$\log g$ retrieval of TRAPPIST-1 with the same priors mentioned in Table~\ref{tab:one_component_priors} after trimming the NIRISS/SOSS spectrum to 1.3--2.4~$\mu$m and masking the same wavelength regions excluded from the SpeX spectra because of strong telluric absorption. 
The same retrieval setup and parameter priors adopted in the main analysis were otherwise used. 
}
{
Figure~\ref{fig:1_corner_plots_wcoverage_test} compares the posterior distributions obtained from the original 1.3--2.8~$\mu$m spectrum and the wavelength-trimmed and masked spectrum matching the wavelength coverage used for the SpeX sample. 
The retrieved $T_1$ (or $T_{eff}$), $R_\star$, $\log g$, and [Fe/H] remain consistent within their uncertainties between the two retrievals. 
In particular, the retrieved surface gravity changes only from
$\log g = 3.995^{+0.222}_{-0.217}$ for the 1.3--2.8~$\mu$m retrieval
adopted in our main analysis to
$\log g = 4.049^{+0.217}_{-0.209}$ when the NIRISS/SOSS spectrum is
restricted to the wavelength coverage used for the SpeX spectra.
This test shows that the TRAPPIST-1 retrieval from NIRISS/SOSS and the retrievals of the other late-M dwarfs from IRTF/SpeX can be reasonably compared and analyzed together, despite the difference in wavelength coverage between the two datasets.
}

\clearpage
\section{Simplified Radiative-transfer Setup for the Activity-associated Toy Heating Experiment}
This appendix summarizes the fixed-VMR prescription used in the simplified radiative-transfer calculation (see Table~\ref{tab:toy_vmr}) and compares the resulting quiet-atmosphere spectrum with the corresponding PHOENIX/NewEra reference spectrum (see Figure~\ref{fig:simple_RT_spectrum_of_quiet_TP}).

\renewcommand{\thetable}{A.\arabic{table}}
\setcounter{table}{0}

\begin{deluxetable}{lcc}
\tablecaption{Adopted Fixed Volume Mixing Ratios for the Simplified Radiative-transfer Calculation
\label{tab:toy_vmr}}
\tablehead{
\colhead{Species} &
\colhead{Opacity Source(s)} &
\colhead{VMR}
}
\startdata
H$_2$        & H$_2$--H$_2$ CIA\tablenotemark{a}
             & $8.4\times10^{-1}$ \\
He           & H$_2$--He CIA\tablenotemark{a}
             & $1.5\times10^{-1}$ \\
H$_2$O       & \citet{Polyansky2018} 
             & $6.7\times10^{-4}$ \\
CO           & \citet{Rothman2010} 
             & $1.0\times10^{-3}$ \\
CH$_4$       & \citet{Yurchenko2024} 
             & $1.0\times10^{-6}$ \\
NH$_3$       & \citet{Coles2019} 
             & $1.0\times10^{-7}$ \\
TiO          & \citet{McKemmish2019} 
             & $2.0\times10^{-7}$ \\
VO           & \citet{McKemmish2016} 
             & $4.0\times10^{-8}$ \\
FeH          & \citet{Wende2010} 
             & $3.0\times10^{-7}$ \\
Na           & \citet{Allard2019a} 
             & $1.0\times10^{-6}$ \\
K            & \citet{Allard2016} 
             & $9.0\times10^{-8}$ \\
$^{48}$Ti    & \citet{Kurucz1995} 
             & $1.0\times10^{-8}$ \\
$^{56}$Fe    & \citet{Kurucz1995} 
             & $1.0\times10^{-7}$ \\
$^{28}$Si    & \citet{Kurucz1995} 
             & $1.0\times10^{-7}$ \\
$^{27}$Al    & \citet{Kurucz1995}
             & $1.0\times10^{-8}$ \\
$^{24}$Mg    & \citet{Kurucz1995}
             & $1.0\times10^{-7}$ \\
H            & H$^-$ continuum \citep{Gray2008} 
             & $1.0\times10^{-4}$ \\
H$^-$        & H$^-$ continuum \citep{Gray2008} 
             & $1.0\times10^{-10}$ \\
e$^-$        & H$^-$ continuum \citep{Gray2008} 
             & $1.0\times10^{-7}$ \\
\enddata
\tablenotetext{a}{The H$_2$--H$_2$ and H$_2$--He collision-induced absorption opacities are from \citet{Borysow1988,Borysow1989a,Borysow1989b,Borysow2001}.}
\tablecomments{
The listed volume mixing ratios are adopted only for the simplified radiative-transfer experiment described in Section~\ref{subsec: A toy moodel to test magnetic heating}. They are vertically uniform and were empirically chosen to provide a reasonable match to the quiet PHOENIX/NewEra reference spectrum over $0.9$--$2.8~\mu{\rm m}$. They should therefore be interpreted as an effective opacity prescription rather than a chemically self-consistent abundance structure for late-M dwarfs.
}
\end{deluxetable}

\renewcommand{\thefigure}{A.\arabic{figure}}
\setcounter{figure}{1}

\begin{figure}
    \centering
    \figurenum{A.3}
    \includegraphics[width=\textwidth, height=0.4\textheight, keepaspectratio]{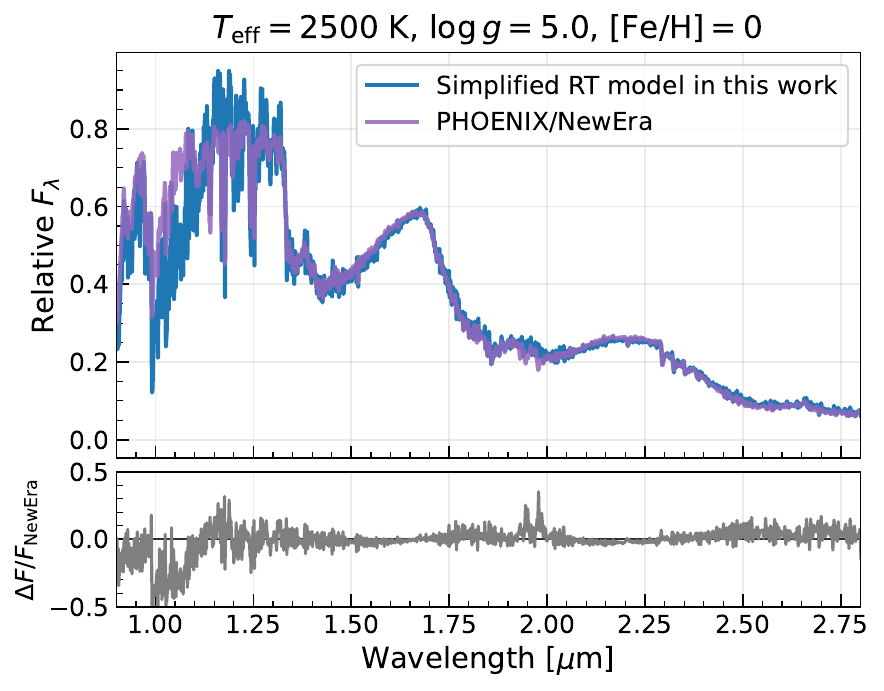}
    \caption{
    Comparison between the simplified radiative-transfer spectrum (in light blue) computed from the quiet T--P profile and the corresponding PHOENIX/NewEra model spectrum (in light purple). 
    The simplified RT spectrum is calculated using \texttt{petitRADTRANS} with the fixed volume mixing ratios of the selected species listed in Table~\ref{tab:toy_vmr}. 
    This setup provides a reasonable match to the corresponding PHOENIX/NewEra spectrum over the wavelength range considered, although noticeable discrepancies remain.
    We therefore use this comparison only as a motivation for adopting the fixed-VMR prescription in our toy modeling, which is designed to test the possible spectral effects of magnetic heating rather than to reproduce a chemically self-consistent late-M-dwarf atmosphere (see Section~\ref{subsec: A toy moodel to test magnetic heating}).
    }
    \label{fig:simple_RT_spectrum_of_quiet_TP}
\end{figure}

%

\clearpage
\vspace{5mm}
\facilities{JWST, IRTF, TESS, Kepler/K2}


\software{astropy \citep{2013A&A...558A..33A,2018AJ....156..123A}, Lightkurve \citep{2018ascl.soft12013L}, \texttt{speclib} \citep{2024AJ....168...82R, speclibv0.1.0b11}, \texttt{petitRADTRANS} \citep[][]{Molliere2019, 2024JOSS....9.7028B}, \texttt{exoTEDRF} \citep{2023Natur.614..670F, 2023MNRAS.524..835R, 2024JOSS....9.6898R}.}





\bibliography{sample631}{}
\bibliographystyle{aasjournalv7.1}



\end{document}